\documentclass[sigconf]{acmart} 
\AtBeginDocument{%
  \providecommand\BibTeX{{%
    \normalfont B\kern-0.5em{\scshape i\kern-0.25em b}\kern-0.8em\TeX}}}

\usepackage{multirow}

\newcommand{\crd}[1]{\textcolor{black}{ #1}}
\newcommand{\rr}[1]{\textcolor{black}{ #1}}

\usepackage{hyperref}
\usepackage{geometry}
\usepackage{enumitem}
\usepackage{array}
\begin{document}

\title[Charting the Future of AI in Project-Based Learning]{Charting the Future of AI in Project-Based Learning:\\ A Co-Design Exploration with Students}


\author{Chengbo Zheng}
\email{cb.zheng@connect.ust.hk}
\orcid{0000-0003-0226-9399}
\affiliation{%
  \institution{The Hong Kong University of Science and Technology}
  \city{Hong Kong}
  \country{China}
}

\author{Kangyu Yuan}
\email{yuanky5@mail2.sysu.edu.cn}
\orcid{0009-0001-8460-9651}
\affiliation{%
  \institution{Sun Yat-sen University}
  \city{Zhuhai}
  \country{China}
}

\author{Bingcan Guo}
\email{bguoac@connect.ust.hk}
\orcid{0000-0002-9001-8727}
\affiliation{%
  \institution{The Hong Kong University of Science and Technology}
  \city{Hong Kong}
  \country{China}
}

\author{Reza Hadi Mogavi}
\email{rhadimog@uwaterloo.ca}
\orcid{0000-0002-4690-2769}
\affiliation{%
  \institution{University of Waterloo}
  \city{Waterloo}
  \country{Canada}
}

\author{Zhenhui Peng}
\email{pengzhh29@mail.sysu.edu.cn}
\orcid{0000-0002-5700-3136}
\affiliation{%
  \institution{Sun Yat-sen University}
  \city{Zhuhai}
  \country{China}
}

\author{Shuai Ma}
\email{shuai.ma@connect.ust.hk}
\orcid{0000-0002-7658-292X}
\affiliation{%
  \institution{The Hong Kong University of Science and Technology}
  \city{Hong Kong}
  \country{China}
}

\author{Xiaojuan Ma}
\email{mxj@cse.ust.hk}
\orcid{0000-0002-9847-7784}
\affiliation{%
  \institution{The Hong Kong University of Science and Technology}
  \city{Hong Kong}
  \country{China}
}

\renewcommand{\shortauthors}{Zheng, et al.}

\begin{abstract}
\crd{
The increasing use of Artificial Intelligence (AI) by students in learning presents new challenges for assessing their learning outcomes in project-based learning (PBL).
}
This paper introduces a co-design study to explore the potential of students' AI usage data as a novel material for PBL assessment. 
We conducted workshops with 18 college students, encouraging them to speculate an alternative world where they could freely employ AI in PBL while needing to report this process to assess their skills and contributions. 
\crd{
Our workshops yielded various scenarios of students' use of AI in PBL and ways of analyzing these uses grounded by students' vision of education goal transformation.
We also found students with different attitudes toward AI exhibited distinct preferences in how to analyze and understand the use of AI.
Based on these findings, we discuss future research opportunities on student-AI interactions and understanding AI-enhanced learning.
}
\end{abstract}





\maketitle

\section{Introduction}
The advance of Artificial Intelligence (AI), especially recent breakthroughs in generative AI (GenAI) and foundation models~\cite{zhou2023comprehensive}, has a foreseeable impact on higher education \cite{neumann2023we, kasneci2023chatgpt, rudolph2023chatgpt}.
This is evident by the increasing use of AI tools by students to assist in their learning tasks~\cite{mogavi2023exploring, chan2023students, lee2022impacts}.
\rr{
Students use AI, such as ChatGPT~\cite{ChatGPT2023}, to resolve confusion and assist with time-consuming tedious tasks, such as debugging and documentation, allowing students to focus more on essential learning tasks~\cite{mogavi2023exploring}.
Despite the benefits of using AI in students' learning, this shift also creates new challenges for education practitioners.}
One critical question that often arises is how to fairly evaluate students' learning outcomes when AI contributes to the completion of learning tasks~\cite{bach2023challenges}.
It is undesirable that assessments end up measuring the capabilities of AI rather than reflecting the students' acquisition and application of skills. 


To tackle this challenge of assessments, some researchers and education practitioners have suggested exercising more in-class or oral tests~\cite{rudolph2023chatgpt}.
\rr{
While this approach may adequately evaluate a student's low-level learning outcomes such as remembering course knowledge, it falls short in measuring students' high-level learning outcomes, such as creative thinking and metacognition, that are highly anticipated by educators in project-based learning (PBL)~\cite{chen2019revisiting, lamb2003project}.
In PBL, students tackle authentic problems and generate artifacts such as reports or models as solutions~\cite{richardson2020we, blumenfeld1991motivating}.
Use of technology in PBL is usually encouraged in PBL~\cite{blumenfeld1991motivating, krajcik2006project}.
Thus, it is conceivable that AI tools will be increasingly adopted by students in PBL with instructors' permission, if not already.
Artifacts produced in PBL usually serve as key indicators of students' learning outcomes~\cite{sterman2023kaleidoscope}. 
However, the increasing use of AI tools in producing these artifacts raises questions about their reliability as accurate measures of student learning \cite{kasneci2023chatgpt, rudolph2023chatgpt, qadir2023engineering}.
}
One alternative strategy is to base the assessment on detailed documentation and reports of the PBL process data, possibly in the form of presentations or learning journals~\cite{blumenfeld1991motivating, sterman2023kaleidoscope}. 
The learning process data can provide insights into key higher-order cognitive processes (e.g., decision-making) that students undergo and qualities (e.g., critical thinking and creative thinking) they exhibit throughout the project. 
Details about how students leverage AI assistance in their projects can be an integral part of future students' learning process data. 
This addition could potentially provide educators with valuable information about the extent to which students' efforts, rather than AI capabilities, contribute to the project outcomes. 
However, a significant gap remains in our understanding of how students might want to report their AI involvement in the PBL process and how such a report could support future assessments of students' learning outcomes in PBL.

In this paper, we aim to take the first step towards filling this gap.
One challenge to our investigation arises from the immature state of AI tool adoption among students.
Despite the popularity of commercial AI products like ChatGPT, many students may lack the necessary skills, such as prompt engineering~\cite{zamfirescu2023johnny, mogavi2023exploring}, to use AI in the way they want.
Besides, many of the AI services are still evolving; they have yet to reach a state of being truly usable and suitable for students in their actual learning activities. 
Even if students have desired AI tools in mind, they may not have adequate access to them due to paywalls, regional restrictions, or concerns about academic integrity when using such aids. 
Nevertheless, the Human-Computer Interaction (HCI) community has a long history of exercising design research methods, such as design probes, co-design workshops, and design fiction, to explore the impact of emerging technologies on different groups of stakeholders and society in futuristic scenarios~\cite{lindley2015back, olson2014ways}.
Inspired by previous design research in the field~\cite{elsden2016metadating, cheon2019beg, luria2023co, wang2022co, wirfs2020giving}, we design and operate a co-design workshop study to engage college students to actively explore the future practices of documenting AI usage in PBL.
The workshop participants are encouraged to speculate a future PBL scenario where they have the freedom to leverage the assistance of any AI capabilities, whether such capabilities currently exist or are yet to be developed, but the assessment to students would largely affected by students' submitted AI usage report, documenting how they have used AI.

Our workshop includes three innovative activities: \textbf{AI-involved PBL journey speculation}, \textbf{Imagine the ideal student}, and \textbf{AI usage report design}.
In the first activity, participants imagined how they might utilize AI in the process if they were to conduct a previous course project again.
This activity echoes the principles of design fiction~\cite{lindley2015back} but situates the speculation not in the distant future but in an ``alternative present''~\cite{cheon2019beg}. 
In the second activity, ``Imagine the ideal student,'' participants envisioned the traits of a future ideal student, such as ``creative'' and ``self-driven'', which they believe should be reflected in their AI-assisted PBL. 
Lastly, the ``AI Usage Report Design'' activity invited participants to craft components of a process report of their re-envisioned course projects specifically related to AI usage, aiming to help with the assessments of the traits (2nd activity) through analyzing students' AI usage behavior (1st activity).

We organized seven separate iterations of a three-hour co-design workshop, with a total of 18 college students, to explore the potential future of reporting AI usage in PBL.
\crd{
We performed qualitative analysis on the collected data and validated our findings using source, investigator, and theory triangulation~\cite{guest2011applied}, ensuring they are rooted in students' experiences and accurately reflect their attitudes.1
}
During our workshop, students produced a broad range of both existing and envisioned AI usages based on their previous learning journeys.
Grounded in these envisioned AI usages, participants suggested multiple methods for analyzing students' interaction with AI, aiming to yield valuable insights for evaluating learning outcomes. 
Post-workshop interviews revealed that students' various attitudes towards AI, led to distinct preferences for how their interaction with AI should be represented in reports. 
However, some participants voiced reservations about the evaluation of their human-AI interactions, citing concerns about the potential for ambiguous interpretation.
\crd{
Due to our study's qualitative nature, statistical generalizability to other scenarios or populations is limited~\cite{smith2018generalizability}. However, it provides detailed insights into students' views on AI-enhanced PBL, encouraging further research in this vital field.
}


\rr{
In summary, this paper contributes to the HCI community by presenting 
1) various future AI usage scenarios, education goal transformations, and possible analysis of students' use of AI in PBL from college students and the nuanced understanding of the reasons behind; 2) A discussion of future research opportunities on student-AI interaction as well as tracking and sensemaking of the students' use of AI based on our workshop findings.
}

\section{Related Work}

\subsection{Project-Based Learning}

Project-based learning (PBL) is a student-centered pedagogy widely adopted in higher education~\cite{kokotsaki2016project}, which stems from the learning theory of active construction~\cite{krajcik2006project, richardson2020we}.
\rr{
Constructivists propose that students learn superficially when receiving information from teachers or computers passively.
In contrast, deeper understanding is achieved when students actively ``construct and reconstruct'' the knowledge through ``experience and interaction in the world''~\cite{krajcik2006project}.
To this end, in PBL, students usually work on a project for an extended period, gaining hands-on experience by creating artifacts, such as reports, models, and videos to answer a driving question~\cite{blumenfeld1991motivating, kokotsaki2016project}.
PBL is also associated with the situated learning theory, which suggests learning would be more effective in authentic contexts~\cite{krajcik2006project}.
Thus, the driving questions in PBL often relate to real-world challenges. 
}
Previous educational research highlights many benefits of PBL, such as better mastery of subject matter~\cite{chen2019revisiting}, promotion of self-regulated learning~\cite{kokotsaki2016project}, sparking students’ motivation~\cite{bell2010project, hira2021motivating}, and improving students’ higher-level cognitive skills such as creative thinking~\cite{chen2019revisiting, guo2020review}.

\rr{
Adopting PBL also presents several challenges, including generating driving questions that are both authentic and relevant to the subject knowledge~\cite{piccolo2023interaction, thomas2000review}; time management~\cite{thomas2000review}; balancing instructor-led guidance and students' self-directed learning~\cite{piccolo2023interaction, kokotsaki2016project}.
Another important challenge is evaluating students learning. 
Education researchers argue that the assessment of PBL should also be ``authentic''~\cite{kokotsaki2016project, bell2010project}.
Traditional tests that can only capture students' low-level understanding of knowledge cannot provide a comprehensive evaluation of students~\cite{blumenfeld1991motivating}.
The artifacts produced by students are frequently used for assessment, but this approach is critiqued for neglecting the process~\cite{sterman2023kaleidoscope, kokotsaki2016project, piccolo2023interaction}.    
As complementary, students are often required to provide in-class presentations, learning journals, portfolios, and self-reflection to show their learning process for assessment~\cite{blumenfeld1991motivating, piccolo2023interaction}.
}

\rr{
This paper explores the impact of AI on future PBL and its implications for student learning assessment.
The integration of technology in PBL is an important research topic~\cite{kokotsaki2016project}.
Technologies are often described as ``cognitive tools''~\cite{thomas2000review}, indicating they help students collect, process, and synthesize information and better engage in higher-order thinking. 
Additionally, technologies empower students to undertake tasks previously beyond their capabilities~\cite{krajcik2006project}, thereby boosting motivation in learning~\cite{blumenfeld1991motivating}.
Besides benefits, \citet{blumenfeld1991motivating} raise concerns about the over-reliance on technology potentially leading to a decline in students' skills and the need to define appropriate roles for teachers and technology.
Previous studies have discussed various technological tools in PBL, such as search engines, project management software, documentation tools and error diagnosis tools~\cite{chen2019revisiting, blumenfeld1991motivating, hira2021motivating, sterman2023kaleidoscope}.
Yet, the impact of AI has been less scrutinized.
Significantly differing from other technologies, AI now demonstrates capabilities that rival or even surpass human intelligence, positioning it as more than just a cognitive tool for students (we will discuss this more in Sec.~\ref{sec:2.2}).
In this paper, we investigate how students might use AI in future PBL, how their learning goals might shift, and how the assessments in PBL should be empowered.
By exploring these questions, we aim to provide insights into designing future PBL instruction and support tools.
}

\subsection{AI Tools to Support Learning Tasks \& Generative AI (GenAI)}
\label{sec:2.2}

\rr{
Extensive research exists on employing AI to support the student learning process.
This includes designing AI to partially replace the teacher's role.
Intelligent tutoring systems, such as those for language and algebra, provide adaptive feedback and problem-solving scaffolding~\cite{weitekamp2020interaction, pardos2023oatutor, peng2023storyfier}.
AI is also deployed to handle class logistics and respond to student inquiries~\cite{wang2021towards}.
Furthermore, there is a growing interest in how AI can collaborate with students during learning.
For example, \citet{jonsson2022cracking} studied students working with a code generation model for creative programming.
They found that errors in AI generation can confuse but also encourage reflection and exploration.
Similarly, \citet{kazemitabaar2023studying} investigated code generation assistance in introductory programming learning, and they found AI could speed up coding without hindering learning.
}

\rr{
While the aforementioned AI tools are designed or selected by teachers, specifically for student learning, in PBL, students have the flexibility to use AI tools not intended for educational purposes.
This include AI tools for brainstorming~\cite{bae2020spinneret}, data science work~\cite{wang2019human}, and creating presentation slides~\cite{zheng2022telling}.
Recently, the rapid development of GenAI, particularly large language models (LLMs), has made AI assistance more accessible to students in the PBL context~\cite{dehouche2023s, mogavi2023exploring}.
LLMs, with their large model scales~\cite{wei2022emergent} and prompting techniques like chain-of-thought~\cite{wei2022chain} and multi-step chaining~\cite{wu2022ai}, demonstrate remarkable proficiency in a wide range of tasks, even achieving success in college-level exams~\cite{chen2023beyond, kung2023performance}.
Moreover, various LLM-based and other GenAI-based tools, such as ChatGPT~\cite{ChatGPT2023} and Midjourney~\cite{midjourney}, are readily available in the market and accessible to students.
According to the content analysis of social media platforms by \citet{mogavi2023exploring}, many college students nowadays have used GenAI tools in their learning, including but not limited to generating review flashcards, creating or editing essays, and assisting peer review.
}

\rr{
Students' use of GenAI in their learning has raised concerns among educators about potential harm.
Generally, it is found challenging to implement responsible AI adoption~\cite{varanasi2023currently, wang2023designing}.
In education, teachers concern that students might use GenAI to cheat on assignments~\cite{lau2023ban, tian2023last}.
This concern is amplified by the challenge in distinguishing between human-written and AI-generated content~\cite{guo2023close}, despite efforts to develop ``AI detectors''~\cite{openai2023edufqa}.
Researchers~\cite{fischer2023generative, rasul2023role, murgia2023chatgpt} and OpenAI's Educator FQAs~\cite{openai2023edufqa} also highlight that GenAI could provide inaccurate, misleading, or biased information, potentially impacting students' learning negatively.
Consequently, some renowned institutions have banned using GenAI tools as an interim solution~\cite{oxford2023, hku2023, cambridge2023}.
}

\rr{
Despite the concerns, education practitioners also widely recognize the benefits of AI and GenAI tools in learning, such as quick, personalized feedback~\cite{mollick2023assigning, hkust2023, rasul2023role}. 
Many foresee a near future where AI usage in student learning becomes a norm~\cite{openai2023edufqa, lau2023ban, rasul2023role}, leading to a transformation in educational assessments, such as emphasizing the learning process rather than just the outcomes~\cite{uw2023, hkust2023} and evaluating students' AI literacy~\cite{hkust2023, long2020ai, rasul2023role}.
People call for more research to explore responsible ways to apply AI in the education field~\cite{murgia2023chatgpt}.
}
\rr{
Moreover, it is widely agreed that transparency in how students use AI is vital for evaluating their learning in the future~\cite{hkust2023, openai2023edufqa, lau2023ban, fischer2023generative}, not only to judge misconduct but also to understand their critical thinking and problem-solving abilities, and to foster students' self-reflection~\cite{fischer2023generative, openai2023edufqa, hkust2023}.
Despite the importance, to our knowledge, there is a lack of research investigating this level of transparency and relevant analysis in an AI-enhanced educational context.
To this end, we investigate a speculative AI-rich PBL setting, focusing on students' needs to collect and analyze their interactions with AI to transparently communicate their AI-enhanced learning process with others.
}

\subsection{Tracking and Sensemaking of Learning Process}

\rr{
Much HCI research has been devoted to studying how to track students' learning process and use techniques, such as learning analytics (LA) dashboards, to help students and instructors understand the tracked data and the learning process.
Much of the learning data is collected from learning-support platforms.
For example, in classroom environments, VisProg~\cite{zhang2023vizprog} collects students' programming data on a Python learning platform and visualizes each student's coding progress to empower instructors to provide in-time feedback;
\citet{yang2023pair} proposed a tool named Pair-Up to track students' learning on digital systems and display students learning status, such as idling and making errors, to teachers to support in-class orchestration.
Collecting video data from remote teaching tools, Glancee~\cite{ma2022glancee} recognizes students' learning status to support instructors' teaching.
In a non-classroom context, students' learning behavior data on learning-support platforms, such as question pool websites, is also studied in previous work to, for example, predict student dropout~\cite{mogavi2021characterizing}, and support student metacognition~\cite{xia2019peerlens}.
}

\rr{
Besides auto-collecting student data from learning-support digital environments, previous research also studies data from students' self-tracking or instructors' observations.
\citet{rong2023understanding} present a qualitative study regarding how Chinese students utilize a data tracking application to self-record various qualitative learning data to support self-directed learning.
~\citet{kharrufa2017group} design Group Spinner, an instructor-facing data tracking and visualization tool.
Instructors can record their observations of students' learning, including the use of technology and outcomes, through Group Spinner, which would then present student data in radar charts to support teachers in the classroom, such as improving communication with students.
In PBL, due to its student-centered nature, students often take the responsibility of tracking their learning data.
For example, \citet{sterman2023kaleidoscope} developed a documentation tool for students in design courses to document their intermediate outcomes in a design project.
Their user studies found that despite the benefits, such as supporting metacognition, students also encounter challenges, such as the tension between ``creation'' and ``documentation''.
}

\rr{
In this paper, we are interested in a learning context that few research has investigated but could become increasingly common in future education.
This involves students learning by doing a project over an extended period of time in non-classroom environments, and AI plays a pivotal role in the learning.
Specifically, students are assisted by powerful AI tools during learning, and their learning goals include ones that might be important in an AI-rich future, such as AI literacy.
While prior works also involve collecting data on students' interactions with AI, their purpose is not to assist education practitioners in understanding the learning process but to answer their unique research questions.
For example, in \citet{kazemitabaar2023studying}'s study about students collaborating with Codex, they collected data on students' AI usage, such as the count of prompts per task and the ``AI-generated code ratio'' (of the final submitted code), to understand whether novices can use AI code generators.
Instead, we study from students' perspectives to explore how their AI interaction data can be leveraged to understand their learning process and evaluate their learning outcomes.
}

\section{Co-Design Workshop Study}

We designed a co-design workshop to investigate the potential of analyzing students' AI usage in PBL for future assessment and invited college students to participate. 

In our investigation, we chose to focus on students rather than instructors for several reasons.
\rr{First, students also play the role of assessors in PBL, including self-assessment~\cite{chen2019revisiting, thomas2000review} and peer assessment~\cite{perez2020project}}.
Second, our study was situated in a future context where AI is both advanced and widely accessible. 
We envisioned that the higher education sector will emphasize responsible usage of AI in learning but cannot constrain how individual students interact with established and emerging AI products and services \cite{mhlanga2023open, qadir2023engineering}. 
In such a scenario, students tend to have a more accurate description of how they may personally leverage AI in PBL than instructors do. 
Third, there are likely other stakeholders in interpreting future AI usage data, such as potential employers, who may evaluate students' qualifications based on their presentation of learning portfolios (e.g., past course projects). 
Students have the agency to analyze their learning data and craft the reporting of their AI usage data in these scenarios.
Lastly, prior research emphasizes the importance of involving students in analyzing their learning data, as they are central stakeholders in their own educational journeys~\cite{alvarez2020deck, sarmiento2022participatory}. 
We carefully considered the alternative of inviting instructors to participate in the co-design workshop alongside students but finally turned down the idea.
For one thing, the inherent power dynamics between instructors and students could impact the latter's design thinking~\cite{schuler1993participatory}.
For another, we sought to include students with diverse PBL experiences for generalizability.
Operational challenges arose in simultaneously recruiting students and instructors whose past PBL activities align closely.
\rr{
In summary, For the purpose of maintaining a focused scope in this paper, we have limited our exploration to students' experiences and perspectives.
Nevertheless, we hope to incorporate teachers’ views in our future work.
}

In the rest of this section, we first elaborate on the participants recruitment and the study setup, and then introduce the three key activities involved in the workshop, which are inspired by previous literature as well as a series of pilot studies.
Lastly, we present the analysis process of the accumulated data. 

\subsection{Participants}

\begin{table*}[]
\caption{An overview of workshop participants' demographics and experience with projects and AI tools}
\resizebox{\textwidth}{33mm}{
\begin{tabular}{cccccccc}
\toprule
WS           & ID & Year      & G & Major                                                                                    & Project Experience                                                                                                & AI Tools Experience                                                                                                            & Freq. of AI Usage \\ \midrule
\uppercase\expandafter{\romannumeral1} & 1  & Senior    & F      & Computer Science                                                                         & Learning system design \& dev.                                                                                    & ChatGPT, Notion AI, New Bing                                                                                                   & Daily                       \\  
                   & 2  & Senior    & F      & Computer Science                                                                         & Booking system design \& dev.                                                                                     & ChatGPT, Github Copilot, New Bing  & Daily                       \\  
                   & 3  & Junior    & F      & Electrical Engineering                                                                   & Charger design \& dev.                                                                                  & ChatGPT                                                                                                                        & Weekly                      \\ 
\uppercase\expandafter{\romannumeral2} & 4  & Sophomore & F      & Literature                                                          & Business analysis of e-vehicles                  & ChatGPT, New Bing, Github Copilot                                  & Daily                       \\  
                   & 5  & Graduate  & F      & Psychology                                                                               & Pedagogy Design                                                                                         & ChatGPT                                                                                                                        & Monthly                     \\  
                   & 6  & Graduate  & M      & Remote Sensing          & Snow Depth Prediction                                                                                         & ChatGPT                                                                                                                        & Weekly                      \\ 
\uppercase\expandafter{\romannumeral3} & 7  & Senior    & M      & Artificial Intelligence                                                                  & Electricity inspection modeling                                                                                           & ChatGPT, New Bing, Github Copilot                                                                                              & Weekly                      \\ 
                   & 8  & Sophomore & M      & Artificial Intelligence                                                                  & Programming project                                                                                              & ChatGPT                                                                                                                        & Monthly                     \\
\uppercase\expandafter{\romannumeral4} & 9  & Graduate  & F      & Management Science                                                       & Rescue Industry Strategy & ChatGPT                                                                                                                        & Weekly                      \\ 
                   & 10 & Senior    & F      & Russian Language                                                                         & Literature project                                                                                     & ChatGPT, New Bing                                                                                                              & Weekly                      \\ 
                   & 11 & Senior    & F      & Arabic Language                                                                          & Study on ancient county records                                                                                           & ChatGPT                                                                                                                        & Monthly                     \\ 
\uppercase\expandafter{\romannumeral5} & 12 & Senior    & F      & Chinese literature                                                          & Hakka songs preservation                      & ChatGPT                                                                                                                        & Weekly                      \\ 
                   & 13 & Senior    & F      & Chinese literature                                                          & Movie casting                             & Notion AI, ChatGPT, New Bing                                                                                                   & Seasonal                    \\ 
                   & 14 & Senior    & M      & Arabic Language                                                                          & Governance analysis \& design                       & ChatGPT, New Bing                                                                                                              & Monthly                     \\ 
\uppercase\expandafter{\romannumeral6} & 15 & Graduate  & F      & Design Strategy                                                           & Market Strategy                                                                                                   & ChatGPT                                                                                                                        & Daily                       \\  
                   & 16 & Graduate  & M      & Interaction design                                                                       & Development of Healing Service                   & ChatGPT                                                                                                                        & Daily                       \\ 
\uppercase\expandafter{\romannumeral7} & 17 & Graduate  & F      &Info. Management  & Oversea purchase service design               & ChatGPT, Notion AI, New Bing                                                                                                   & Daily                       \\ 
                   & 18 & Senior    & F      & Architecture                                                                             & Store Leads Design                                                                                                & ChatGPT, Notion AI                                                                                                             & Weekly                      \\ \bottomrule
\end{tabular}
}

\label{table:participant}
\end{table*}

We recruited participants by disseminating recruitment messages with registration forms through various channels, including social media, word-of-mouth, and posters at four higher education institutions in East Asia. 
Following this, we received 68 applications for the workshops.
We carefully screened the applications and filtered candidates with inadequate experience in PBL.
\rr{
For instance, we received five applications from first-year undergraduate students. 
However, the projects they described, such as building a personal web page or learning a programming language, were not solving real-world problems. 
Thus we considered they lacked PBL experience and did not include them in the workshops.
}
Additionally, we also required participants to have experience in using AI tools.

We recruited 18 participants (female=13, male=5) from diverse backgrounds.
\rr{
Our qualitative study's sample size was determined by reaching theoretical saturation~\cite{guest2011applied}. 
This was evidenced by no new insights emerging from the last two workshops, indicating a sufficient data breadth for our research objectives.
}
Our participants consisted of both undergraduates (12) and graduate students (6), and their majors varied from Computer Science (2), Engineering (2), Data Science (4), Design (3), Psychology (1), Literature (3), and Language (3). 
In terms of AI tool experience, all participants interacted with ChatGPT, while a subset (8) also used other AI tools, such as Notion AI. 
Most participants were frequent users of AI tools, with 7 using them weekly and 6 using them daily.
\rr{
When asked to self-assess their ability to apply AI to solve real-world problems on a 5-point Likert scale, participants' average score was 3.44 (SD=0.90).
Additionally, 14 participants reported having utilized AI in their recent course projects.
For instance, P07 used ChatGPT to ``explore the pros and cons of various neural network designs,' while P04 used ChatGPT for ``data analysis tasks because it is good at dealing with data''.
}
The detailed demographic information of our participants is presented in Table.~\ref{table:participant}.
Each participant was compensated with \$10 per hour for their participation.

\subsection{Workshop Setup}
\label{sec:setup}
Our workshops took place in the summer of 2023 and allowed participants to join in person or virtually.
In-person attendance was restricted to three cities where institutions at which we had disseminated recruitment messages are based. 
However, due to challenges arising from the disparate geographical locations of the participants in each available time slot, all workshop sessions eventually took place virtually.
We utilized a video conferencing software~\footnote{\url{https://voovmeeting.com/}} as well as a virtual collaboration whiteboard tool, \textit{Miro}~\footnote{\url{https://miro.com/}}, to conduct all co-design workshops.
To ensure the smooth execution of these workshops, at least one day prior to each session, we sent out necessary materials such as the consent form, workshop guidance, and the link to the \textit{Miro} whiteboard, that was going to be used in that workshop session, to the participants. 
The workshop guidance included instructions on pre-workshop preparations (detailed in Sec.~\ref{activity1}) and a brief guide to using the \textit{Miro} platform.

Our pilot studies suggested that our workshop generally lasted around 3 hours.
Each workshop was divided into two parts on the same day to mitigate potential fatigue and optimize engagement. 
Our workshops had three key activities (see Sec.~\ref{sec:workshop_design}) and the first part, which took place in the late afternoon and lasted for two hours, covers \textit{activity 1}, \textit{activity 2}, and the first step of \textit{activity 3}. 
Participants reconvened for the second part in the evening, which lasted an hour and was dedicated to completing the second step of \textit{activity 3}.


\subsection{Workshop Design}
\label{sec:workshop_design}

\begin{figure*}
    \centering
    \includegraphics[width=0.9\linewidth]{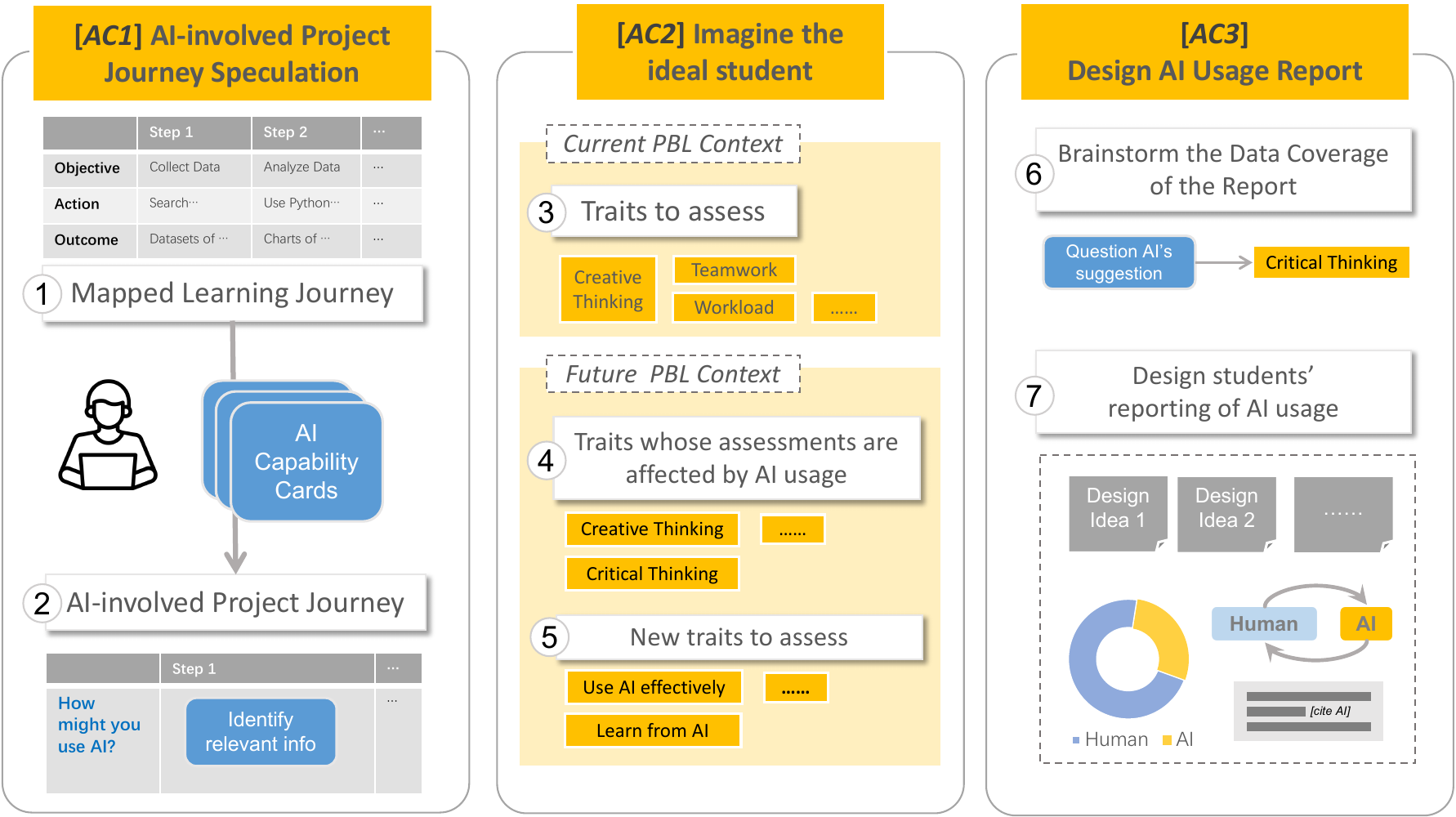}
    \caption{\rr{
    Our co-design workshop involved three activities (AC). In AC1, students mapped their learning journey (1) and envisioned AI integration in projects with the help of AI Capability Cards (2). AC2 involved identifying key traits for student assessment in a PBL context (3), AI's impact on these assessments (4), and emergent traits necessary in an AI-rich future (5). In the last activity (AC3), based on the outcomes of the prior two activities, students first considered what data should be covered in a report on their AI usage in PBL (6) and then visually designed the report (6).
    }}
    \label{fig:workshop-flow}
\end{figure*}

Our workshop design was finalized through four rounds of pilot workshop studies.
The supplementary material provides our design and reflection of each pilot study.

Our overarching vision informs the framework of our final workshop.
That is, in a future where AI plays a significant role in student learning, both the learning process and the focus of assessments will undergo considerable transformation.
Under this assumption, our workshop first engaged participants in picturing how AI might influence their learning processes (\textit{activity 1}), then explored potential changes in the assessments of PBL due to the integration of AI in learning (\textit{activity 2}).
We posit that materials for measuring student performance will also need to be updated in response to assessment transformation. 
Students' AI usage could serve as a valuable data addition to this evolving assessment landscape. 
Hence, in our workshop, grounded in the insights from \textit{activities 1 \& 2}, participants are encouraged to conceptualize their own methods for analyzing and reporting AI usage for assessment purposes (\textit{activity 3}).
Figure~\ref{fig:workshop-flow} illustrates all the activities in our workshop.
In the following subsections, we detail the design and the techniques we used in each activity.

\begin{figure*}
    \centering
    \includegraphics[width=0.97\linewidth]{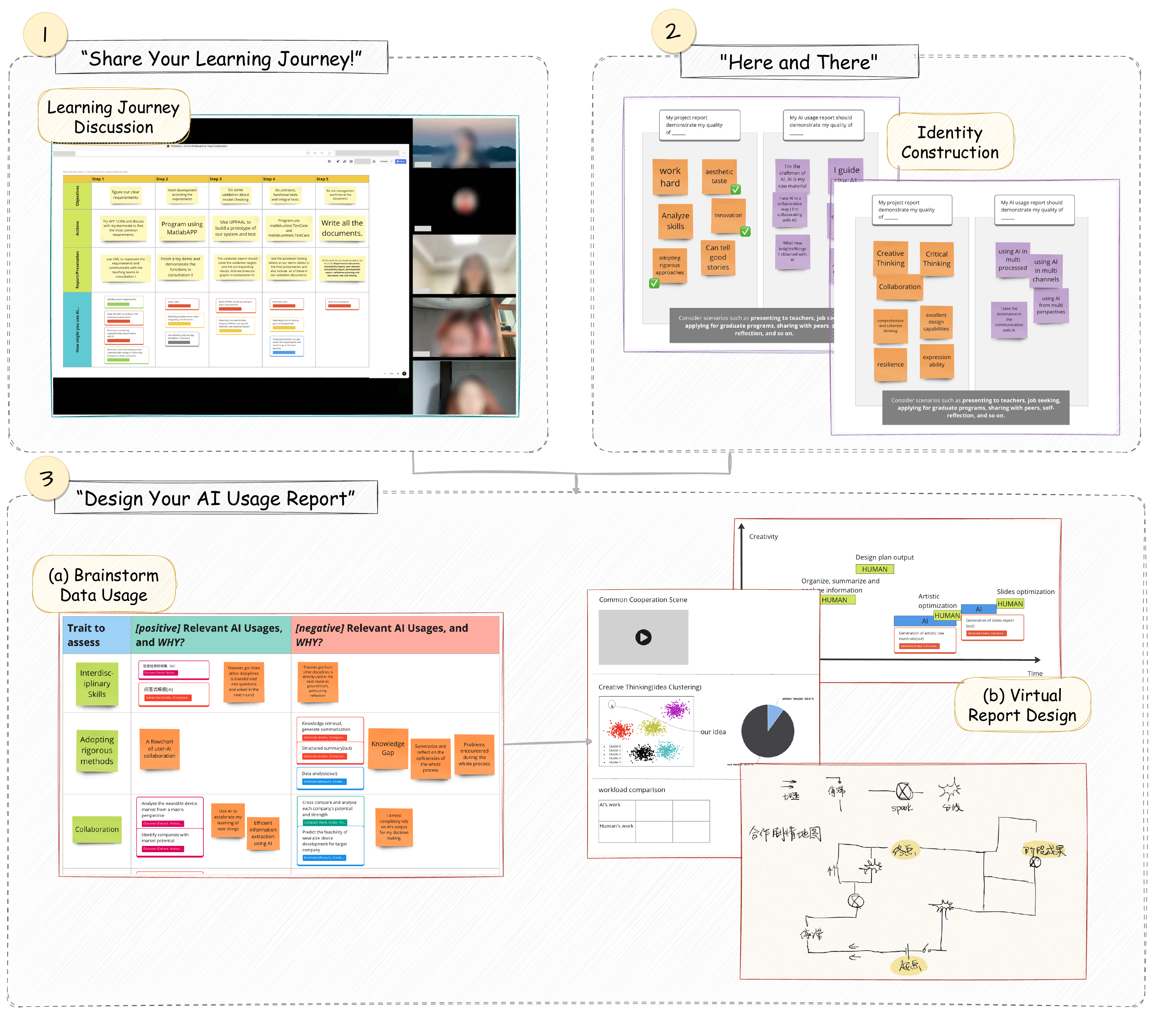}
    \caption{\rr{The workshop utilized the Miro platform for collaborative activities. The facilitator shared the Miro board screen throughout (1). In the first activity, students described their project journey in our provided learning journey table (1). The second activity involved identifying current and future PBL traits using colored sticky notes—orange for current PBL importance, purple for future needs, and orange with ticks for traits affected by AI's rise (2). In the third activity, students linked traits from the second activity to AI usage, using sticky notes for detailed notes (3(a)). For the visual report, they could use Miro's tools for design mockups (e.g., P16), incorporate external materials (e.g., P02), or sketch using preferred tools (3(b)).}}
    \label{fig:miro-flow}
\end{figure*}

\subsubsection{Activity 1: AI-involved Project Journey Speculation}
\label{activity1}
In this activity, we encouraged participants to envision future AI capabilities they might leverage in PBL freely. 
Initially, we planned to employ the Futuristic Autobiographies (FABs) method often used in design fiction studies~\cite{cheon2018futuristic, cheon2019beg, wirfs2020giving}.
Specifically, we provided a PBL context that included one project topic given by researchers (e.g., ``lung cancer prediction'') and a futuristic education context (e.g., ``teachers encourage AI use and any AI tools you want are ready'').
However, this approach fell short in our pilot studies; participants struggled to generate concrete ideas about the needs and challenges they might face in the hypothetical project, which in turn limited their creative thoughts about AI use.

To overcome this limitation, we adopted the concept of the ``alternative present'' from design fiction literature~\cite{auger2010alternative, cheon2019beg} in the final workshop design.
Instead of imagining a future project, participants were asked to recall a recent or memorable PBL experience they actually had. 
They were then prompted to re-imagine these projects in a world where AI technology is ten years more advanced than today. 
This approach allowed participants to ground their speculations in concrete past experiences, enhancing the richness and feasibility of their envisioned AI applications.

Operationally, after signing the consent form, participants were asked to revisit and document a memorable or recent PBL experience using our learner journey template prior to the workshop.
\textbf{Mapping learner journey} is a co-design technique that captures and communicates the essential phases and overarching flow of a student's learning experience~\cite{prieto2018mapping, prieto2018co}. 
The template prompted them to detail each step of their PBL process, from objectives to actions and specific outcomes. 
Participants were allowed to adapt the template to better fit their individual experiences (Fig.\ref{fig:workshop-flow} (1)). 

At the beginning of the workshop, we first introduced the background of our study, asking all participants to introduce themselves, and then invited participants to share their mapped-out learner journeys. 
Then, they were tasked with a 15-minute brainstorming session, envisioning how AI technology -- presumed to be ten years more advanced -- could augment their past PBL processes.
We used the ``AI capability cards'' presented by ~\citet{yildirim2023creating} to foster creative thinking as props.
These cards categorized AI functionalities into eight types~\footnote{Including ``Estimate'', ``Forecast'', ``Compare'', ``Detect'', ``Identify'', ``Discover'', ``Generate'' and ``Act''~\cite{yildirim2023creating}.}.
Participants were introduced to each AI capability category with examples under education contexts.
They could play and customize the AI capability cards at will and put down new AI capabilities not captured by the existing categories on ``wild cards'' (Fig.~\ref{fig:miro-flow} (1)).

After \textbf{integrating proposed AI usage into their learning journey maps}, participants took turns sharing their revised project journey (Fig.\ref{fig:workshop-flow} (2)).
We encouraged the audience to actively contribute additional AI ideas for each other's projects. 
Participants took a 10-minute break before going into \textit{activity 2}.

\subsubsection{Activity 2: Imagine the Ideal Student}

After exploring the potential uses of AI in PBL, the next objective of our workshop was to identify future challenges and needs related to assessing students' PBL performance. 
This would inform the subsequent design of AI usage reports tailored to these assessments.
Specifically, this activity involved two key steps: identifying traits that are challenging to assess due to AI incorporation and postulating future traits of students that may emerge with more prevalent AI usage in PBL (Fig.~\ref{fig:workshop-flow} (4, 5)).
Here, we use the term ``trait'' to refer to the qualifications and qualities a student could exhibit from their learning processes, such as critical thinking and creativity, and might be desired by instructors and others to assess.

\textbf{Identifying traits whose assessments are challenged by AI incorporation:} 
In the first step, participants reviewed the assessment goals of their past PBL experiences recollected in \textit{Activity 1}.
These goals could be self-defined, considering the autonomous nature of PBL \cite{blumenfeld1991motivating, krajcik2006project}, or set by external stakeholders like instructors. 
Participants reflected on what traits were deemed essential in those assessments (Fig.~\ref{fig:workshop-flow} (3)) and how to measure them in the past. 
Subsequently, they were asked to critically examine how their AI usage, as speculated in \textit{Activity 1}, might affect the assessment of these traits.

\textbf{Imagining future traits needed: } 
In the second step, we asked participants to take 10 minutes to brainstorm new traits, such as the ability to apply diverse AI tools in disciplinary tasks, that might become desirable when AI plays a pivotal role in both society and education.  
This aligns with literature suggesting that technological advancements can reshape educational paradigms~\cite{romrell2014samr}.

To facilitate this speculative thinking, we employed an adapted persona-building technique commonly used in participatory design to analyze user needs~\cite{sarmiento2022participatory}.
Unlike traditional persona-building, which is rooted in past experiences, our revised approach asks participants to envision ``ideal future students'' and their traits related to AI experiences. 
This encouraged participants to think beyond existing educational frameworks and consider emerging needs and challenges.

After brainstorming the traits of an ``ideal student,'' we extended the discussion by asking participants to take 5 minutes to consider how these traits might be valued differently depending on assessor identity: instructors, the students themselves for self-assessment, and future employers. 
This exercise enabled a comprehensive understanding of what learning outcomes students would want to present to different stakeholders in an AI-aided PBL environment. 
At the end of \textit{activity 2}, participants were invited to discuss the traits they had thought of and also commented on others' ideas.

\subsubsection{Activity 3: AI Usage Report Design}
In this activity, participants designed their desired report of AI usage to better understand their learning behavior in PBL.
The report should surface their interaction with AI (\textit{activity 1}) and aim to align with the traits they value (\textit{activity 2}).
This activity was divided into two steps: brainstorming what AI usage data needs to be covered in the report and visually crafting an AI usage report.

\textbf{Brainstorm the Data Coverage of an AI Usage Report:} We asked each participant to choose one to two traits from \textit{activity 2} and produced ideas about what interactions with AI that appeared in their speculated AI-aided PBL (\textit{activity 1}) could be relevant to these traits (Fig.~\ref{fig:workshop-flow}(6)).
The goal was to prompt participants' reflection on what aspects of students-AI interactions in PBL may be worth being analyzed, curated, and presented in their AI usage report for learning assessment purposes.
Besides, we would like participants to recap their works in the first two activities so that their latter design can be grounded in their previous thoughts.
In particular, participants were suggested to map any speculated AI usages from \textit{activity 1} to the relevant traits brought up in \textit{activity 2}. 
To facilitate individual brainstorming (15 minutes), we provided several example types of interaction data about one specific AI usage: the types of AI, at what stages in the project this AI usage occurs, the input to AI, any customization to the AI tools, and how the AI output is handled.
Participants used sticky notes to add comments on how each type of interaction data about a specific AI usage can be aligned or misaligned with the trait(s) they want to project (Fig.~\ref{fig:miro-flow}).
They were encouraged to consider not only their own selected traits but also those proposed by other participants.
Once done, they took turns sharing their results and giving feedback to other students. 

\textbf{Visually Depicting an AI Usage Report:}
In the second step of \textit{activity 3}, we described the concept of a hypothetical ``\textit{magic project studio}'', which could catch all students' interactions with AI and generate an AI usage report based on students' needs.
Participants were given 30 minutes to design AI usage reports that they desired this ``magic project studio'' to create (Fig.~\ref{fig:workshop-flow}(7)).
Using this concept, we hope participants bypass implementation details and focus on brainstorming \textit{what} can be presented in such a report to fully explore the potential of students' AI usage data as an assessment material.
During the design process, we encouraged participants to iterate their outcomes in prior steps, such as brainstorming the data coverage of the AI usage report (\textit{activity 3} step 1), whenever a new idea came to their minds.

As introduced in Sec.~\ref{sec:setup}, this step was conducted in the evening of the workshop.
Before introducing this step, we took 5 minutes to recap the content of the first part of the workshop. 
Then, participants were instructed to consider how the reports should be framed in the scenario of submitting them to their course instructors.
They were also encouraged to consider how such reports can be modified in other contexts, including self-reflection and job seeking.
We provided visual components, including charts, tables, and dialogue bubbles, in the Miro whiteboard as design material.
We also encouraged participants to use any other methods to showcase their design ideas, such as sketching or text descriptions. 

Lastly, each participant was asked to share their report designs.
Besides introducing the report design itself, they were tasked with demonstrating what traits they believed their designs could be used to assess, how the chosen human-AI interaction data are related to their assessment goal, and any other design ideas they had in mind but had difficulties illustrating.
After participants of one workshop session had all presented their designs, they were then suggested to comment on others' proposals, including the pros and cons, and have an immediate discussion among themselves.

After participants completed \textit{activity 3}, we also conducted a 15-minute follow-up focus group interview with them, including questions including ``\textit{what do you think of the importance of leverage students' AI usage data in future education}'', ``\textit{from a student's perspective, how would you like the AI usage report, such as the one you finally designed, be produced and delivered to others, e.g., your instructor?}'', and ``\textit{what are your general experience of the workshop, any confusing moments?}''

\subsection{Data Analysis}
We recorded all workshop sessions, accumulating approximately 24 hours of audio footage.
These recordings were initially auto-transcribed and subsequently manually verified for accuracy. 
We employed the Inductive Thematic Analysis method to analyze the data~\cite{guest2011applied}.
\crd{
The inductive approach offers flexibility in uncovering the nuances of the data, which is particularly beneficial in studies like ours that explore relatively uncharted territories~\cite{guest2011applied, braun2006using}.
}
Three researchers -- including a facilitator and an assistant who participated in all workshops -- engaged in the data analysis.
They first familiarized themselves with the data by reviewing the recordings several times.
Then, the coders independently coded the transcripts.
They met frequently during the analysis to discuss any discrepancies. 

\rr{
    We integrated a dynamic approach in analysis, simultaneously analyzing early workshops while running later ones.
    This concurrent analysis allowed us to triangulate data effectively.
    We asked participants in later workshops about their views on findings from earlier ones, enriching our understanding and validating our results. 
    This method, coupled with cross-referencing the data with Miro boards used in each workshop (as source triangulation), ensured a robust, iterative analysis.
}

\rr{
After completing the analysis, we adopted member checking to validate our findings, which involved inviting participants to review our findings and assess their alignment with their intentions and experiences~\cite{guest2011applied}.
Specifically, we followed the synthesized member checking~\cite{birt2016member}.
Our preliminary analysis was summarized in a concise five-page report. This document encapsulated the main themes, essential codes, and representative quotes. We approached participants for their assistance in reviewing this report, and 13 consented. 
These participants were provided with the report, and we requested them to annotate and provide feedback on any aspects they found either reflective of or inconsistent with their intentions and experiences.
Of the participants, 11 returned the annotated reports.
We carefully compared the feedback from these reports with our existing codes. 
This comparison allowed us to refine our analysis, ensuring it reflected the participants' perspectives and experiences more accurately.
}

\section{Findings}

In this section, we present six primary themes that emerged from our analysis. The first three themes are aligned with the three activities conducted during our workshop, while the latter three themes surfaced from participants' comments and discussions.

\subsection{Speculated AI Usages in PBL by Students}
\label{sec:AI-usages}

Our seven workshops with 18 participants in total resulted in over 100 student-desired AI usages.
Through our analysis, six subthemes regarding the purposes of these usages emerged.

\begin{itemize}
    \item \textit{Automating Repetitive and Time-consuming Tasks.}
All students desire AI to improve process efficiency by automating activities perceived as time-consuming, monotonous, and laborious, such as collecting data, documenting the implementation, and debugging.
    \item \textit{Supporting Divergent Thinking.}
Students hope AI can stimulate creative and out-of-the-box thinking by providing diverse ideas (``\textit{AI has randomness, and the results it comes up with each time may be different, and I may let it give me strategy ideas multiple times for more perspectives}'') and filtering ideas (``\textit{Let AI exclude published and commercially available related application ideas}'' (P1)).
    \item \textit{Supporting Selection from Alternatives.}
AI is expected to aid students in choosing the most effective ideas through analysis and comparison. For example, P03 wanted AI to compare his ideas of algorithms and P05 would like AI to compare different literature to extract the ``\textit{most correct conclusion}'' to use in her project.
    \item \textit{Drafting or Direct Implementing of Solutions.}
Students hope AI can take their solution idea from conception to realization. 
This would involve coding or creating prototypes based on the student's foundational concept. 
For example, P18 said, ``\textit{I might express my ideas to the AI after I've formed a solution for myself and let it do this last step of visualization for me.}''
    \item \textit{Feedback on Solutions.}
This involves AI evaluating the solution's effectiveness and offering suggestions for improvement.
Participants mentioned 15 times that they hoped AI could help them evaluate the proposed solution based on its effectiveness, feasibility, rigorousness, etc.
    \item \textit{Guiding Students to Learn.}
This involves AI’s educational capabilities, from teaching new concepts to evaluating students' knowledge readiness to do the project. 
\end{itemize}

\subsection{Students' Envisioned Future Assessment Transformation}
\label{assess-transform}
Participants have diverse and sometimes conflicting ideas when considering how their speculated AI usage might impact existing assessment methods.
Participants also developed novel traits they believed were needed for an ideal future student.
We introduce these visions from students, which serve as interpreters of the purposes of students' analysis of AI usage data introduced in the next section.

\subsubsection{Old Traits Made Different by AI}
Most participants felt that traditional assessment methods, such as those based on artifacts, are inadequate for evaluating traits like creative thinking and efficiency, given the generative capabilities of AI. 
However, some participants (P02, P06, P14, P16) argued that artifacts should still hold significant weight in assessments.
This belief was rooted in their assumptions about AI's limitations, such as its ability to offer only coarse-grained analyses and its inability to tailor solutions to a specific project context.

Interestingly, while most participants considered logical and critical thinking vital skills, P04 and P16 commented that these skills might not be critical in the future, given AI's growing reasoning capabilities.
P17 opposed this view, stating: ``\textit{How you choose to talk to the AI, what you ask it to clarify or expand on—that all takes some serious critical thinking}.''



\subsubsection{New Traits Needed Due to Students' AI Adoption}
\label{new-traits}
These traits are those not necessarily accessed currently but are believed by participants to be very important due to AI usage.

\begin{itemize}
    \item \textbf{Efficacy in using AI.} This trait is the most frequently mentioned trait, through which participants highlight that an ideal student should use AI with clear purposes (P06, P10, P16), clearly communicate intents to AI (P11, P13, P15), and as a result, the output from AI lead to efficiency or performance increase.
    \item \textbf{Leadership in Project Direction.} Participants envisioned the ideal student as someone who retains control over the project's direction, relegating AI to a ``supporting actor'' who executes tasks as directed by the human leader.
    \item \textbf{Symbiotic Learning between Students and AI.} Some participants (P04, P07, P08, P16, P18) appreciated the notion of a mutually beneficial relationship between humans and AI: AI contributes valuable knowledge or capabilities, while students, in turn, refine AI functionalities to suit the project's needs better.
    \item \textbf{Judgment and Discernment.} Participants, including those with design backgrounds (P15, P16, P18), argued that traditional design skills may become less critical as AI becomes powerful in designing. Instead, the ability to judge quality and make wise selections from alternatives could be essential.
\end{itemize}





\subsection{Students' Designs of Reporting of AI Usages}
\label{4.3-students'design of reporting of AI usages}
\begin{figure*}
    \centering
    \includegraphics[width=\linewidth]{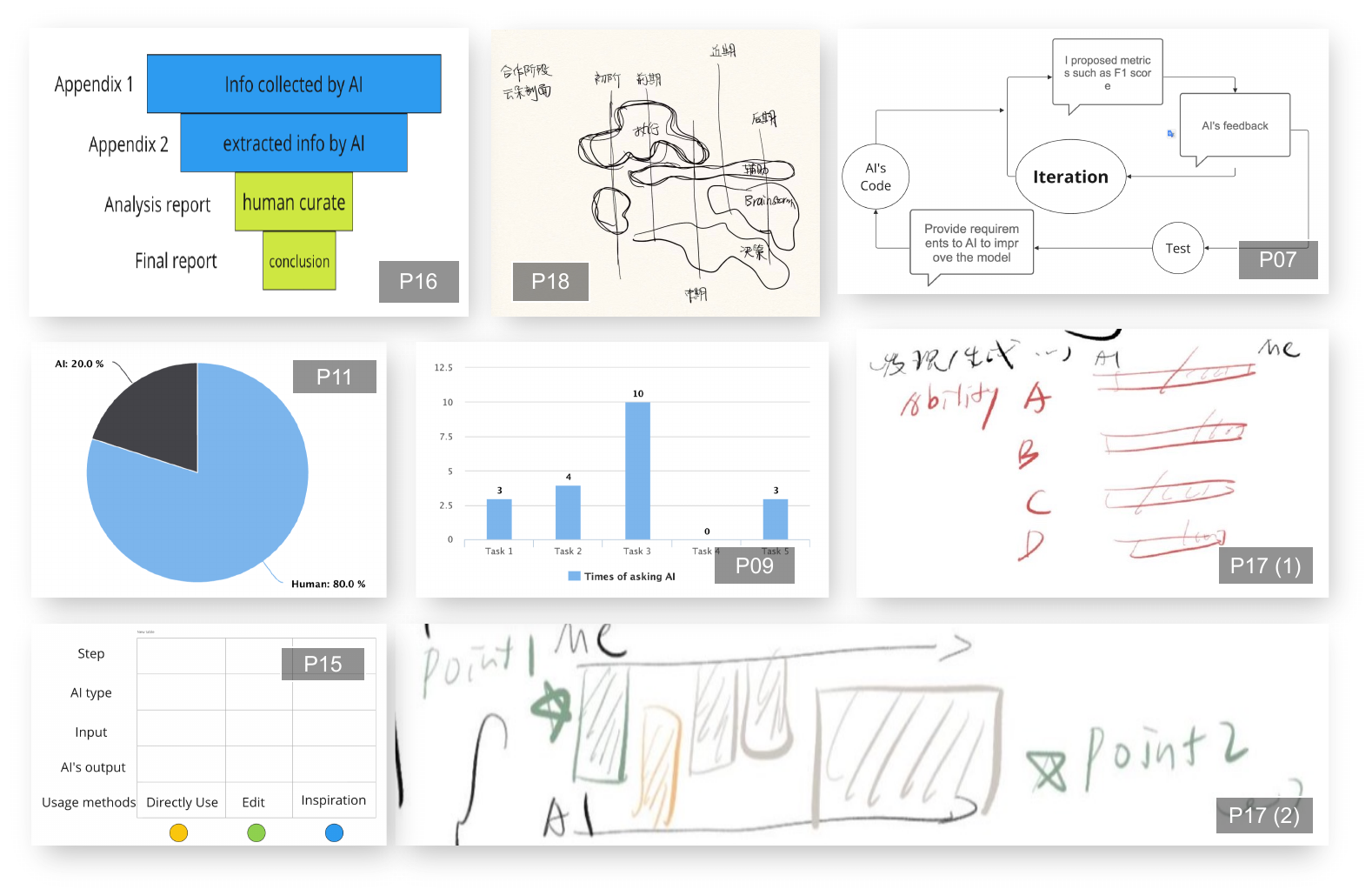}
    \caption{\rr{A subset of designs of the reporting of students' AI usage resulted from step 2 of \textit{activity 3} of the co-design workshop. We explain the designs when we mention them in the main text.}}
    \label{fig:gallery}
\end{figure*}

This section introduces the participants' proposed ideas for mining their AI usage data.
In the \textit{activity 3}, participants brainstormed what data insights could be extracted from the AI usage data, which we refer to as \textbf{usage analysis}; and also tried to visually depict these insights, which we refer to as \textit{framing idea};.
Nevertheless, not every usage analysis has corresponding data framing, possibly limited by participants' data and visualization literacy, but it is worth future research.

In the following subsections, we introduce the themes of usage analysis we discovered.
For each theme, if applicable, we discuss the key \textbf{usage analysis} underneath the theme with associated \textit{framing ideas}.
We elaborate on why participants came up with this usage analysis, which is tightly related to their imagined AI usages and envisioned future assessment transformation.
Note that we do not claim that the resulting categories of usage analysis and data framing are exclusive or representative.
Instead, we aim to open up the discussion space of the potential value of analyzing students' AI usage data through these categories.

\begin{table*}[h!]
\centering
\begin{tabular}{@{}m{0.2\linewidth}m{0.05\linewidth}m{0.72\linewidth}@{}}
\toprule
\textbf{Themes} & \textbf{Freq.} & \textbf{Description of key codes under the themes} \\
\midrule
Task allocation between students and AI & W=9, M=9 & 
\begin{itemize}[leftmargin=*]
    \item[\textendash] Learning gains differ across tasks; Students should work on tasks with higher learning gains.
    \item[\textendash] Humans and AI have different relative advantages; Each party should work on the tasks that suit them best.
    \item[\textendash] It is difficult to allocate tasks clearly in real human-AI teaming.
\end{itemize} \\
\cline{1-3}
\addlinespace
Quantifying and Depicting AI’s engagements & W=14, M=8 & 
\begin{itemize}[leftmargin=*]
    \item[\textendash] Compute a percentage of AI’s contribution and students’ contribution
    \begin{itemize}[leftmargin=*]
        \item[\textendash] High percentage of student contribution can be used to support artifact-based assessment.
        \item[\textendash] High percentage of AI contribution can show students’ efficacy in using AI.
        \item[\textendash] It is difficult to define the computing method.
    \end{itemize}
    \item[\textendash] Connect the students’ AI usage to the final artifacts to support evaluation based on artifacts.
    \item[\textendash] Categorize how AI exactly engages in the project process, such as sparking ideas or causing conflicts.
\end{itemize} \\
\cline{1-3}
\addlinespace
Effectiveness of Students-AI Interaction & W=14, M=8 & 
\begin{itemize}[leftmargin=*]
    \item[\textendash] Effectiveness of students’ inquiries to AI in getting desired assistance.
    \item[\textendash] The evolvement of AI behavior due to students’ involvement.
    \item[\textendash] Outcome improvements due to students’ AI usage.
\end{itemize} \\
\cline{1-3}
\addlinespace
The process of students incorporating AI’s suggestions into the project & W=14, M=11 & 
\begin{itemize}[leftmargin=*]
    \item[\textendash] Students’ subjective reflection on how they treat AI’s suggestions.
    \item[\textendash] Present the process of students filtering, editing, and re-questioning AI’s suggestions.
    \item[\textendash] Present the diversity of opinions considered when making decisions with suggestions from AI.
    \item[\textendash] Discover the iteration of student-AI interaction, showing whether students and AI build on each other's work.
\end{itemize} \\
\cline{1-3}
\addlinespace
Quantifying Students development through human-AI interaction & W=6, M=9 & 
\begin{itemize}[leftmargin=*]
    \item[\textendash] Discover student's behavior changes in interacting with AI to show whether students develop their skills in using AI throughout the process.
    \item[\textendash] Quantify students' learning based on how they have delivered tasks to AI.
\end{itemize} \\
\cline{1-3}
\addlinespace
AI impact on student-student collaboration & W=3, M=4 & 
\begin{itemize}[leftmargin=*]
    \item[\textendash] Comparing student-student interaction with student-AI interaction to show whether AI usage negatively impacts students' collaboration.
    \item[\textendash] Analyze whether team members’ attitudes toward AI cause conflicts within teams.
\end{itemize} \\
\cline{1-3}
\addlinespace
Students’ ethical awareness in using AI & W=5, M=8 & 
\begin{itemize}[leftmargin=*]
    \item[\textendash] Analyzing whether students’ AI usage obeys regulations and respects people's privacy.
\end{itemize} \\
\bottomrule
\end{tabular}
\caption{Summary of key codes under various themes of students' designs of AI usage analysis. The frequency column presents how frequently the themes were mentioned by participants in the workshops (W) and resonated in the member checking (M).}
\label{table:1}
\end{table*}

\subsubsection{Task Allocation between Students and AI}
\label{task-allocation}
Many students wish to \textbf{differentiate tasks handled by humans and those executed by AI} in their reports.
An important design consideration stems from the students' view that different tasks within the project have different weights in nurturing and showcasing various skills.
Therefore, presenting who -- human or AI -- conducted specific tasks implicitly indicates skill development or mastery. 
This notion is exemplified in P16's task allocation framing, depicted in Fig.~\ref{fig:miro-flow}.
Using a \textit{Gantt chart-style visualization}, P16 illustrates the division of tasks between humans and AI, represented by differently colored blocks with task annotations. 
The chart also includes time spent on each task (x-axis) and the level of creativity required (y-axis).
P16 rationalized this design by stating, ``\textit{The project is a learning journey, and key to that learning is the execution of tasks that cultivate specific skills like creative thinking}.''

One very different consideration is that humans and AI have advantages in different tasks. 
Thus for complementary performance, they want to allocate specific tasks to the party that is good at them.
For example, P08 liked to use a \textit{flow chart} to show that he uses AI mainly for \textit{Automating Repetitive and Time-consuming Tasks}, as he believed the advantage of AI is to perform these tasks fast; P01 would like to show the time she spent on tasks that can be easily and quickly done by AI such as debugging.
On the other hand, P10 would like to show that students themselves are handling creative tasks, although without concrete framing, and mentioned that ``\textit{When creative ideas are needed, I don't think AI is helpful even if it lists a lot of data out and make a perfect analysis. Because the spark of inspiration needs a particular moment.}''.

It is interesting to know that although some students would like to show the task allocation, in their real action, they did not want to really clearly allocate the tasks, which often originated from the belief that they and AI were working as a team.
P16 designed a table that clearly shows what AI did and what she did, but she stated that:

\begin{quote}
    Although AI is responsible for a certain part, my input is involved. As a team, we do need to divide so clearly. [...]
    I'm not using this chart to illustrate my specific actions but to show it to the teacher to easily assess my abilities.
\end{quote}

\subsubsection{Engagement of AI in the Project}
\label{4.3.2-engagement of ai in the project}
Besides differentiating human and AI labor discretely at task-level, many participants would like to quantify to what extent AI has engaged with humans in the project.
The most common usage analysis proposed by students was to \textbf{analyze the percentages of AI's contribution versus students' in the project}, often framed as \textit{pie chart} (e.g., Fig.~\ref{fig:gallery}, P11), in either project-level or detailed task-level.
The motivation of such an analysis is sometimes to complement the artifact-based assessments, under which students expect the chart showing their contribution to be much larger than AI; thus, the artifacts can represent the students' own skills.
But sometimes, the motivation is about to show students' \textit{efficacy in using AI}, and participants (P02, P16) expect the work done by AI to be much larger than humans, showing they could leverage AI to assist them in many pieces of stuff.

\rr{
Participants suggested various measurements of AI's contribution, including AI-generated word count (P04), AI usage frequency in projects (P14), and the number of problems AI solved (P18). 
However, considering the complex human-AI interaction, some (P04, P16, P17) questioned the objectivity of these metrics. 
P17 pointed out that even if 80\% of a report is AI-generated, the human contribution should be valued more, considering significant human input to AI and testing with AI.
}

While such percentages could be too abstract to understand and hard to measure, some students proposed to \textbf{connect AI usages to the artifacts} to more concretely represent the AI engagement, to complement further the assessment based on artifacts.
Participants suggested using \textit{the citation-reference style} to annotate any places AI has made a difference.
Moreover, P15 proposed to \textbf{detail the types of AI engagement} (Fig.~\ref{fig:gallery}, P15), including directly using AI output, student-edited AI output, and human content inspired by AI.
While with a similar idea of differentiating different kinds of AI engagement, P18 expected to annotate the AI engagement based on the timeline of the project (Fig.~\ref{fig:miro-flow} (P18)), which mainly aimed for assessing students' \textit{efficacy in using AI}.
She differentiated four types of AI engagement: AI facilitates the project process, AI stalls the progress, AI sparks the idea, and AI causes conflicts.


\subsubsection{Effectiveness of Students-AI Interaction}
\label{student-mastery}
Usage analysis under this theme mainly aims to assess whether a student exhibits the trait of ``efficacy of using AI'' introduced in Sec.~\ref{new-traits}.
Students consider \textbf{whether their question designed could well prompt AI to get desired assistance} works as an important indicator of their mastery of using AI.
P11 wants to present her question design process as well as the final question; P02 would like to have a \textit{video recording} of her series of questions to AI highlighting her skills, for example, how her scaffolding the questions ``\textit{I give it a broad requirement and then see if it generates a good result and if not, I refine the question step by step.}''
Students (P11, P17) also consider metrics such as \textit{the number of questions and time needed in questioning and answering with AI} as indicators of whether students could effectively use AI, and they believed that fewer questions and less time in questioning AI for one specific task indicate better AI usage skills.
Besides, P02, P09, and P18 also want to compare their AI usage data with their peers to show their mastery.

Several students considered presenting \textbf{the changes of the AI's behavior due to students' continuous input to AI} to show whether they successfully guided AI towards the direction they like to show mastery, although without concrete data framing ideas.
P13 expressed that :
\begin{quote}
    Let's say at the beginning, AI is just a basic general AI, but I feed it some papers, and it gradually understands the stuff that I might be trying to do, and then it can give some matching help.
\end{quote}

\textbf{Another usage analysis is mining the project outcome improvements due to students' AI adoption}, which is often framed using the comparison between students' original work with the work improved by AI.
This usage analysis is not only used to showcase students' AI mastery but also to foster students' reflection to improve their AI mastery.
For example, P17 suggested that:

\begin{quote}
    I'd like to mark some points of the conversation where asking questions or keywords in the back-and-forth dialogue [between AI and me] made it possible to progress with our project or make a breakthrough. [...] It would tell me how to talk to AI better, which is valuable.
\end{quote}

\subsubsection{The Process of Students Incorporating AI’s Suggestions into the Project}
\label{decision-making}

Students hypothesized that using AI could bolster their decision-making capabilities by supporting divergent thinking, selecting alternatives, and providing feedback on solutions.
However, many students knew that AI might provide incorrect information, introduce bias, and misguide their decision-making; thus, students must prevent adopting AI's ideas without caution.
Such concerns could turn into opportunities for assessing students.
Students believe examining their behavior in incorporating AI's suggestions can offer key insights for assessing traits such as critical thinking, creativity, and leadership in human-AI interactions.

One straightforward approach to understanding how students incorporate AI-generated ideas involves asking them to \textbf{articulate their perceptions and reflections on the AI's suggestions}. 
Beyond this subjective analysis, several participants expressed a desire to \textbf{demonstrate their process of filtering, editing, and questioning AI's suggestions}.
For example, P16 created a funnel chart (Fig.\ref{fig:gallery}, P16) to visualize how AI-generated insights undergo multiple layers of scrutiny. 
P17 depicted a diagram that showcases the bidirectional information exchange between humans and AI (Fig.\ref{fig:gallery}, P17(2)). 
Although not visually represented, P17 expressed interest in tracing which AI-provided inputs progressed to subsequent stages and their ultimate impact. 
Similarly, P07 constructed a bar chart to display the frequency with which she questioned AI's suggestions, asserting that a higher frequency of questioning indicated more critical thinking (Fig.~\ref{fig:gallery}, P09).

Moreover, some students emphasized their wish to \textbf{highlight the diversity of opinions considered when incorporating AI's suggestions}. 
For example, P14 stated, ``\textit{I want to show that I am synthesizing multiple AI's suggestions. For example, I use ChatGPT for initial ideas and then turn to New Bing for additional perspectives.}''

\rr{
Finally, some students emphasized the need to display \textbf{the iterative process of blending AI suggestions with student inputs}. 
This demonstrates mutual enhancement in projects. 
For example, P07 used a flowchart (Fig.~\ref{fig:gallery}, P07) to show how students and AI collaboratively refine a model, with neither party's ideas being used without the other's feedback.
}

\subsubsection{Quantifying Student Development through Human-AI Interaction}
\label{student-develop}
\rr{
Our participants thought that data from student-AI interactions could offer a valuable understanding of how students' skills evolve throughout the project. 
One aspect examined is \textbf{the development of student's skills in using AI}.
P02 expressed a desire to demonstrate how she got useful assistance from AI through step-by-step inquiries, illustrating her gradual mastery of effective AI usage.
}

Moreover, some students were aware that relying too heavily on AI for specific tasks could potentially hinder their skill development; as a result, they desire metrics to \textbf{quantify such effects}.
P17 used a bar chart to capture the accumulated negative impact of adopting AI in the project across time (Fig.~\ref{fig:gallery}, 17(1)).
She explained that ``\textit{there would be scores for skills such as creative thinking, and whenever the student chooses to complete some tasks using AI, there would be some deduction [to the scores].}''
P11 designed a similar chart, but instead of a deduction, she would like the score to increase whenever the student did something manually or had rich interaction with AI, such as many follow-up questions, on a certain task.

\subsubsection{AI Impact on Student-Student Collaboration}
\label{sec:stu-collaboration}

PBL often involves teamwork, and several participants indicated that the integration of AI might affect collaboration, warranting assessment.
For example, P03 and P18 suggested that \textbf{the ease of communicating with AI might discourage students from actively communicating with human teammates}, which might be inappropriate for students to practice their collaboration skills.
\rr{In member checking, three other participants found this point resonated with their experience.
P04 added that ``\textit{I prefer asking AI for assistance first, then share the results with teammates for discussion.}''
}
P03 and P18 recommended an analysis that contrasts the frequency and quality of student-student communication against student-AI communication.

Additionally, P01 suspected \textbf{divergent attitudes toward AI within teams would result in conflicts}, which should be identified in the analysis.
\rr{
In member checking, P04 and P16 indicated they experienced such conflicts in their projects.
P16 mentioned that:
\begin{quote}
We generally agree to use AI for topic selection and framework building. However, some team members disagree with using AI to generate content due to quality and integrity concerns.
\end{quote}
}
Despite the need for analysis, participants did not develop a framing idea for exposing the AI impact on collaboration, which is worth future research. 

\subsubsection{Students' Ethical Awareness in Using AI}

A few participants suggested whether students used AI responsibly was worth analysis, for example, obeying regulations (P12) and respecting people's privacy (P10).
The framing idea was mainly posting documentation of the AI students use.
P16 described a framing idea: ``\textit{Suppose I used AI to draw a picture, but the AI's training data that support its drawing were from several painters, and it would be nice to have a tree diagram of the source of this intellectual property.}''

\subsection{Different Envisioned Roles of AI}
\label{4.4-envisioned_roles}
We noticed distinct differences among participants regarding their design goals and final reporting frameworks for analyzing their AI usage.
Upon analyzing their rationales during the workshop and the member check results, we identified three students' beliefs on the role of AI, each of which significantly influenced how participants analyzed and framed AI usage data:

\begin{itemize}
    \item \textbf{AI as a tool.} Some participants viewed AI as a mechanism to augment human abilities. 
    Statements like ``\textit{AI should not replace humans in execution}'' (P05, P14) and ``\textit{AI should only handle trivial tasks}'' (P09) were shared among this group. 
    These individuals were generally interested in highlighting their ``\textit{leadership in directing the project,}'' often through lower levels of AI engagement.
    \item \textbf{AI as a teammate. } Another group of participants (e.g., P02, P04, P13, P18) saw AI more as a collaborator. For them, the overarching goal was to complete the project effectively as a team. 
    As such, they questioned the necessity of separating human traits from AI interaction and considered the final artifacts of the project to be weighted much more than considering the student-AI interaction process. 
    P13 commented: ``\textit{I think a good human-AI relationship should involve a blended, mutual engagement, so differentiating our work from AI's may not be necessary or desirable.}''
    \item \textbf{AI as an Expert.} A third group (e.g., P11, P16) saw AI more as an expert resource they could consult, albeit one whose advice could be subjective, biased, or misleading. P11 noted,
    \begin{quote}
    When AI becomes almost perfect, it develops its own 'thoughts' or 'goals.' [...] As a result, I could end up losing my original focus.
    \end{quote}
    For these participants, traits like critical thinking were essential. 
    They believed reports must assess how cautiously students integrated suggestions from these AI experts.
\end{itemize}

\subsection{Impacts of Scenarios}

In \textit{Activities 2 \& 3} of our workshop, we encouraged participants to reflect on how AI usage might be analyzed across three contexts: instructor assessment, job-seeking evaluations, and self-assessment. 
Participants generally advocated for a holistic, in-depth analysis of AI usage for instructor assessments to inform learning assessments. 
In contrast, when considering job-seeking, the emphasis shifted towards showcasing efficiency in leveraging AI technologies. 

For self-assessment, the focus generally turned to empowering reflection. 
P16 categorized his AI interactions based on the purposes of facilitating learning or merely serving project goals.
He exported to the former ones in his self-assessment report. 
P07, meanwhile, advocated for integrating AI usage data with personal metrics like emotions and heart rates, arguing that this would enrich reflective practices, which aligns with previous research on fostering self-reflection~\cite{li2011personal, rong2023understanding}.

\subsection{Concerns of Reporting Students AI Usage to Enable Assessment}
\label{concern}
Most participants acknowledged the value of analyzing students' interaction with AI for assessment purposes.
However, two concerns were raised.

First, participants were concerned about \textbf{the fairness of such evaluation adds-on}.
Based on prior experiences with GenAI tools, participants pointed out that students faced difficulties critically evaluating suggestions from powerful AI. 
P07 noted that students might not be ``\textit{thoughtlessly accepting AI's suggestions},'' but could be settling due to these suggestions being good enough and lack of better alternatives.
However, if instructors only rely on student-AI interactions for assessment, it may result in unfairly low scores for students regarding critical thinking under the assumption that students over-rely on AI.
P17 raised an additional concern that modifying AI-generated content could be mistakenly attributed to a student's critical and creative thinking. 
The modification only reflects ``\textit{the student's external, contextual knowledge that the AI lacks}.''
The inherent limitations in AI's sensing and understanding could inadvertently lead to unwarranted accolades for students without careful inspection.

\rr{
Second, some participants (P04, P17, P18) considered \textbf{students might ``hack'' to get a ``beautiful'' report of AI usage}.
P17 mentioned that students were likely to change their learning behavior to cater to the better AI usage report recognized by instructors.
}

\section{Discussion}

Our workshops provided valuable insights into students' use of AI in future Project-Based Learning (PBL). 
\rr{
We found various ways students might use AI and, from the student's perspective, the potential learning goal shifts.
Our participants generally believed whether students can effectively use AI would be an important future assessment criterion.
More importantly, participants suggested that the student-AI interaction data can not only be used to augment traditional assessments by approaches such as linking project artifacts to specific AI usage but also offer a window into their higher-order thinking skills and skills in effectively using AI.
However, our analysis also revealed nuances, such as varied student beliefs about AI's role in learning, which in turn influence their engagement with the technology. 
Students also raised practical concerns regarding analyzing students' use of AI to understand student learning, including fairness and the potential for hacking behavior by students.
In this section, by triangulating these findings with existing literature, we identify new research opportunities in student-AI interaction and tracking and sensemaking of students' use of AI for education and HCI researchers. 
This section also discusses the generalizability of our results, the limitations of our study, and our future work.
}

\rr{
\subsection{Research Opportunities on Student-AI Interaction}
}

\begin{table*}[h]
    \centering
    \begin{tabular}{@{}p{0.35\linewidth}p{0.3\linewidth}p{0.3\linewidth}@{}}
        \toprule
        \textbf{Research Opportunity} & \textbf{Relevant Results} & \textbf{Relevant Literature} \\
        \midrule
        \hyperref[sec:dis-AI-role]{How do students' perceptions of AI roles influence educational interactions?} & Different beliefs on AI roles affecting use and analysis (Sec.\ref{4.4-envisioned_roles}); mismatches in beliefs may lead to conflicts (Sec.\ref{sec:stu-collaboration}). & AI roles in HCI research~\cite{zheng2022telling, kim2023one, jakesch2019ai, zheng2023competent}. \\
        \cline{1-3}
        \addlinespace

        \hyperref[sec:dis-tailor]{How can we tailor AI for students to use in PBL?} & 
        Concerns about powerful general-purpose AI (Sec.\ref{concern}).
        & Domain-specific GenAIs~\cite{wu2023bloomberggpt, zhang2023huatuogpt}; Need to adapt AI for educational usage~\cite{murgia2023chatgpt}. \\
        \cline{1-3}
        \addlinespace
        
        \hyperref[sec:dis-srl]{How can we support self-regulated learning in AI-enhanced environments?} & 
        Designs for examining students' self-regulation (Sec.\ref{concern}). & Self-regulated learning~\cite{zimmerman2002becoming}; Tools to support SRL~\cite{rong2023understanding, sterman2023kaleidoscope}. \\
        \cline{1-3}
        \addlinespace
        
        \hyperref[sec:dis-goal-balancing]{How should education practitioners balance the goals of effective use of AI and actively learning in future PBL?} & 
        Students wanted to delegate tasks better done by AI (Sec.\ref{task-allocation}).
        & PBL and active learning~\cite{barron1998doing, blumenfeld1991motivating}; task delegation in human-AI interaction~\cite{lai2022human, shi2023retrolens}. \\
        \cline{1-3}
        \addlinespace
        
        \hyperref[sec:dis-communication]{What are the impacts of AI on communication in education?} & AI is replacing instructor roles (Sec.\ref{sec:AI-usages}).
        & Role of instructors in PBL~\cite{kokotsaki2016project, blumenfeld1991motivating, krajcik2006project}. \\
        \bottomrule
    \end{tabular}
    \caption{Summary of research opportunities on student-AI interaction, derived from our study results and previous literature.}
    \label{tab:ROStudentAI}
\end{table*}

\rr{
    \subsubsection{How do students' perceptions of AI roles influence educational interactions?}
    \label{sec:dis-AI-role}
    Our research revealed a diversity of opinions among participants regarding the roles AI should assume, ranging from a \textit{tool} to a \textit{teammate} or an \textit{expert}. 
    These roles significantly influence their conjectures on AI utilization and the subsequent analysis of such use.
    Previous HCI research has explored various potential roles for AI, including those of an ``assistant''~\cite{zheng2022telling, kim2023one}, ``mediator,''~\cite{kim2023one, jakesch2019ai} or ``equal decision-maker''~\cite{zheng2023competent}.
    However, the discussion focuses primarily on the implications of the \textit{designers' framing of AI roles} for end-users.
    With the evolution of AI towards serving more general purposes~\cite{wei2022emergent}, \textit{users} have much more autonomy in using AI in their desired way.
    Our findings suggest that users' beliefs about what roles AI should play also matter, which deserves future research on the broader impacts.
    For example, in the educational contexts examined in this study, mismatched beliefs about AI's role between students or between students and teachers may create conflicts or result in ineffective pedagogical designs.
}

\rr{
    \subsubsection{How can we tailor AI for students to use in PBL?}
    \label{sec:dis-tailor}
    In our workshops, we encouraged students to envision utilizing any AI tools in PBL. 
    However, another potential future learning environment involves students using AI that is specifically fine-tuned for education purposes, suggested by the development of domain-specific GenAIs~\cite{wu2023bloomberggpt, zhang2023huatuogpt}.
    Our study findings reveal potential friction when students use powerful general-purpose AI and suggest directions for fine-tuning future student-facing AI tools for PBL.
    Participants expressed concerns that powerful AI threatens the fairness of assessments based on student-AI interaction data, since students might have limited judgment abilities regarding AI's outputs and may merely accept AI's outputs without question.
    Our findings echo the call for adapting AI for educational usage~\cite{murgia2023chatgpt}.
    Future work can explore a more student-centered design of AI.
    For example, designing personalized student-facing AI tools that align with their capabilities or creating AI systems that scaffold responses based on the student's skill levels, offering guidance or direct assistance as appropriate.
}

\rr{
    \subsubsection{How can we support self-regulated learning in AI-enhanced environments?}
    \label{sec:dis-srl}
    Self-regulated learning (SRL), which is defined as learners actively controlling their learning process~\cite{zimmerman2002becoming}, is integral to PBL and other problem-based learning activities~\cite{kokotsaki2016project, zimmerman2003motivating}. 
    While AI tools might offer valuable feedback, there's a risk that students' over-reliance on AI could impede critical SRL steps such as self-assessment and the independent adjustment of learning strategies~\cite{zimmerman2002becoming}. 
    Acknowledging this, our study participants suggested emphasizing the analysis of how students incorporate AI's outputs to assess whether students are using AI inappropriately (Sec.\ref{decision-making}). 
    They also suggest monitoring students' interactions with AI (Sec.\ref{student-develop}), which is relevant to the self-monitoring concept in SRL~\cite{zimmerman2002becoming}. 
    Similar to previous research~\cite{rong2023understanding, sterman2023kaleidoscope}, our participants' design aims to promote documentation and learning analysis practices to support SRL.
    Future research should empirically examine the impacts of AI-enhanced environments on SRL and investigate the effects of documentation and learning analytics on students' AI reliance and autonomy in learning.
}

    \subsubsection{How should education practitioners balance the goals of effective use of AI and actively learning in future PBL?}
    \label{sec:dis-goal-balancing}
    PBL engages students in solving real-world problems.
    But there is a risk that students may fall into a situation where the ``doing'' of a project takes precedence over ``doing with understanding''~\cite{barron1998doing}.
    In previous PBL, these two goals have had the potential to complement each other, as succeeding in practical tasks generally requires students to develop specific skill sets.
    However, the advent of AI technologies adds a layer of complexity.
\rr{
    Many participants considered practicing and demonstrating skills in effectively utilizing AI important for future PBL (Sec.~\ref{assess-transform}).
    They considered tasks that AI can do better should be delegated to AI.
    These ideas echo previous research on effective human-AI collaboration in the workplace~\cite{zheng2022telling, shi2023retrolens, lai2022human}. 
    In this way, the growing capability of AI suggests students would be in an oversight position for many tasks in PBL, including some that require creativity and critical thinking, which will help students understand knowledge better.
    However, PBL's foundation lies in constructive learning theories, where students learn through active engagement~\cite{blumenfeld1991motivating, krajcik2006project}. 
    The task delegation to AI can bypass these critical active learning steps.
    To this end, the goal of effective use of AI could harm students' active learning.
    Future research should investigate how to balance these two goals.
    One opportunity is to instruct students to use AI in a way that they can actively construct knowledge.
    For example, many participants mentioned students should spend time carefully crafting and guiding AI to get effective assistance from AI.
    Future research can study whether, in the input crafting and engagement process, students can ``construct and reconstruct'' knowledge mentally and actively learn from the process.
}

\rr{
    \subsubsection{What are the impacts of AI on communication in education?}
    \label{sec:dis-communication}
    Some participants wanted to use AI to partially, if not totally, replace the instructors' position in PBL, such as \textit{providing feedback on solutions} and \textit{guiding students to learn}.
    Such AI usage might not be appropriate as although PBL is student-centered, instructors still play a significant role in it~\cite{kokotsaki2016project}.
    Without adequate student-instructor communication, students might learn in a direction that does not match the curriculum and instructors' teaching plans.
    Besides, AI could also impact student-student communication (Sec.~\ref{sec:stu-collaboration}).
    Future research should consider more comprehensively examining the effects of AI on educational communications, especially with longitudinal study design.
}

\subsection{Research Opportunities on Tracking and Sensemaking Students Use of AI}
\begin{table*}[h]
    \centering
    \begin{tabular}{@{}p{0.4\linewidth}p{0.25\linewidth}p{0.3\linewidth}@{}}
        \toprule
        \textbf{Research Opportunity} & \textbf{Relevant Results} & \textbf{Relevant Literature} \\
        \midrule
        \hyperref[sec:dis-data-collection]{How can we support the collection of data around students' use of AI?} & 
        Student-proposed various analysis needs (Sec.\ref{4.3-students'design of reporting of AI usages}).
        & HCI research on documentation tools~\cite{sterman2023kaleidoscope, rong2023understanding}; information provenance~\cite{lindley2018exploring, han2022passages}. \\
        \cline{1-3}
        \addlinespace
        \hyperref[sec:dis-AI-contribution]{How can we make sense of AI's contribution based on students-AI interaction data?} &
        Difficulty in separating human-AI contributions (Sec.\ref{4.3.2-engagement of ai in the project}).
        & Human-AI symbiosis~\cite{licklider1960man} \\
        \cline{1-3}
        \addlinespace
        \hyperref[sec:dis-multiple-perspectives]{How can we support sensemaking of students' use of AI from multiple perspectives?} & 
        Different interpretations of AI usage (Sec.\ref{4.3-students'design of reporting of AI usages}).
        & ``One chart, many meanings''~\cite{ahn2019designing}. \\
        \cline{1-3}
        \addlinespace
        \hyperref[sec:dis-motivation-documentation]{How can we motivate students to document their use of AI and report it honestly?} &
        Potential hacking behavior of students (Sec.\ref{concern}).
        & Benefits communication and reflection nudges~\cite{xia2020using}. \\
        \bottomrule
    \end{tabular}
    \caption{Research opportunities on tracking and sensemaking students' use of AI in learning.}
    \label{tab:ROStudentAIDocumentation}
\end{table*}

\subsubsection{How can we support the collection of data around students use of AI?}
    The first step to analyzing students’ AI usage is to collect relevant data.
    While the interaction log of students and AI serves as the most direct data, our findings provide insights into several other types of data worth collecting from students' AI usage, including:
    \begin{itemize}
        \item \textit{Contexts when using AI}. 
        Our study shows the analysis needs to differ based on when and why students use AI.
        For example, when AI is used for automating tasks, the focus would be on the types of learning tasks managed by AI and student proficiency with AI.
        For using AI for feedback on solutions, participants expect to examine the detailed process of how students incorporate AI suggestions.
        \item \textit{Students' thoughts and actions with AI's suggestions}. The analysis theme favored by our participants, ``the process of incorporating AI's ideas into the project,'' requires examination of students' thoughts and actions.
        \item \textit{The lineage from students' AI usage to their solutions}. Solutions that students come up with, such as artifacts in PBL, are still considered important assessment materials. Connecting students' AI usage to the corresponding parts of the solution might help education practitioners understand students' contributions.
    \end{itemize}
    \rr{
        It is non-trivial to collect these data.
        The first two types of data might need input from students.
        Previous HCI research studies how students document their artifact-based learning data~\cite{sterman2023kaleidoscope} or multi-modal learning data~\cite{rong2023understanding} and how interactive tools might help with the documentation.
        Future research can look into how students document their motivation and thoughts when using AI during learning, investigate what specific challenges students can encounter, and what designs of documentation tools can be helpful.
        Moreover, research on information provenance through interactions~\cite{lindley2018exploring, han2022passages} can provide insights for collecting the third data type.
        For example, future research can explore how to reify the transformation from AI outputs to solutions.
    }

\rr{
\subsubsection{How can we make sense of AI's contribution based on students-AI interaction data?}
Students considered making sense of AI's contribution to the project essential.
Some participants provided various ideas on the computing methods of AI's contribution, but others suspected that the evolving complex human-AI interaction would make it difficult to disentangle the contributions of two parties.
These findings echo an early discussion on human-AI symbiosis.
\citet{licklider1960man} conjectures it is difficult to separate the contribution of humans and AI in decision-making.
But \citet{licklider1960man} also mentions that, overall, humans should provide leading contributions by doing tasks such as goal setting and judgments.
Future research should further explore signals of whether students are in the leading position when collaborating with AI.
We believe the signals should not necessarily be single values, such as percentages, as many participants imagined (e.g., Fig.~\ref{fig:gallery}, P11).
One might study how to gather qualitative and quantitative evidence from student-AI interaction data on whether students are leading their projects compared to AI and invite education practitioners to engage in sensemaking of students' and AI's contribution more comprehensively. 
}

\rr{
\subsubsection{How can we support sensemaking of students' use of AI from multiple perspectives?}
Our study provides insights into the diverse lenses one can adopt to analyze students' use of AI.
}
Moreover, our study reveals intriguing complexities regarding the values students attach to using AI, which significantly impact the sensemaking of the analysis results.
The diversity of values aligns with and amplifies the ``one chart, many meanings'' consideration in learning analytics~\cite{ahn2019designing}.

For example, the analysis of question-and-answer rounds and time spent communicating with AI serves divergent purposes for different student groups.
One faction sees fewer rounds and shorter time as evidence of students' efficient mastery over AI. 
Conversely, another group interprets more rounds and longer time as indicative of a careful, critical engagement with AI's suggestions.
Likewise, students understand the pie charts showing AI's impact on project results differently.
Some participants, like P09 and P12, aim for a more minor AI contribution arc to highlight their significant human-led efforts. 
Others, such as P02 and P16, aspire to demonstrate a larger AI contribution to showcase their ability to leverage AI capabilities fully. 

Another nuanced example is found in the analysis of task allocation between students and AI (see Sec.~\ref{task-allocation}).
A group of students aims to analyze whether focusing on certain tasks leads to better learning.
Another group uses the data to show that humans and AI are suited for different tasks. 
As a result, while both groups agree on using AI for repetitive tasks and humans for creative and decision-making roles, the first group values this for its educational benefits, and the second sees it as practical due to AI's current limits.
However, as AI evolves to become more personalized, context-sensitive, and creative, the perspectives of the second group suggest that roles involving critical thinking, decision-making, and creativity may increasingly be transferred to AI (P04 and P16 already have such a tendency), which conflict with the educational ideals of the first group.

In the above cases, we do not seek to discuss which values are more ``correct'' or beneficial.
\rr{
However, such diversity underscores the need for education practitioners to interpret students' use of AI carefully.
Future research should examine the interpretation space and how to support fair and comprehensive sensemaking of students' use of AI.
For example, one may study how to involve students themselves in the interpretation better, considering students' self-assessments are always considered essential for successful PBL~\cite{thomas2000review}.
}

\rr{
    \subsubsection{How can we motivate students to document their use of AI and report it honestly?}
    Our participants admitted that if their use of AI is considered one of the ways to assess their learning, they will probably hack up a nice report of AI usage to get a higher grade.
    This matches with teachers' expectation that students might not tell how they use AI honestly~\cite{lau2023ban}.
    Future research can study in what ways we can motivate students to document and report their use of AI faithfully.
    Literature provides some potential directions.
    First, education practitioners might leverage various methods to communicate the benefits of faithful AI usage documentation and report to students.
    \citet{xia2020using} proposed to use visualization to nudge students to reflect on their behavior of ``gaming the system''.
    In our case, one might communicate with students how their AI usage might negatively impact their learning, how documentation might balance that, and how an honest report can help instructors provide better instructions.
    Second, the assessment of students should not only be based on students' reports of their AI usage.
    Instructors might emphasize that such a report is used to understand students' learning, and the final assessment would be made by synthesizing multiple factors.
    Overall, we propose that the documentation of students' use of AI should be framed as helping students better learn instead of as a grading tool to motivate them in documentation and reporting.
}



\subsection{Limitation \& Future Work}
\rr{
This paper presents a qualitative investigation into the potential future of students' use of AI in PBL based on workshops with 18 college students.
Our participants are from four East Asian institutions with diverse major backgrounds.
Given the qualitative nature of our study, we cannot assert that our findings generalize to a broad range of scenarios (e.g., PBL in courses not engaged by our participants) or to a larger population (e.g., students from other institutions) in a statistical-probabilistic sense, nor can we ensure their applicability over extended periods~\cite{smith2018generalizability}.
Nevertheless, the qualitative approach of this paper lets us dive into a growing important learning scenario (i.e., AI-enhanced PBL and its assessment) due to the rapid development of AI, provide in-depth insights into students' beliefs and needs, and motivate relevant future research.
Future research could build upon our work by quantitatively examining our findings, including the effectiveness of various analytic designs, students' anticipated roles for AI, and the influence of scenarios on students' needs, using larger student samples and more extended study periods.
}

Our study also presents several additional limitations.
First, the format of our investigation is limited to 3-hour workshops, while the PBL usually extends over weeks or even months. 
While we prompted participants to draw upon their prior long-term PBL experiences for our activities, a longitudinal study involving actual PBL settings is a promising next step.
Such a study could yield deeper insights into how students would like to interact with AI, and analyze and present AI usage data.
Second, the co-design activities in our study were based on hypothetical AI usage, driven by our aim for generalizability in light of rapidly advancing technology. 
However, hands-on experience with AI in PBL is invaluable for generating more nuanced perspectives on how AI can be leveraged. 
As an extension to our current work, we envision encouraging students to employ existing AI tools in the aforementioned long-term study while speculating on desired future capabilities.
Lastly, our workshops primarily focused on eliciting student perspectives.
Incorporating the viewpoints of educators by exposing the findings of our workshops could provide a more comprehensive understanding and assessment of students' AI usage suggestions.

\section{Conclusion}
\rr{
In conclusion, this paper presented a co-design study exploring the potential of utilizing students' AI usage data to understand student learning in project-based learning (PBL).
The study provided insights into the opportunities and challenges of analyzing students' AI usage data. 
Participants envisioned how they would use AI in future PBL and highlighted the impact of AI on assessment transformation. 
They proposed various designs to analyze students' use of AI to examine students' skills, decision-making processes, and ethical awareness in using AI. 
We also found different students have different beliefs in the role AI should play in their projects, from a tool that augments their abilities to a teammate or expert.
Such belief impacts how they want to use AI and report their AI usage.
This research contributes to the HCI community by offering insights into future practices related to AI usage in education and informing the design of AI education systems, project documentation tools, and learning analytics systems. 
It advances our understanding of how AI can shape student learning and assessment in PBL contexts.
}
\begin{acks}
This work is supported by the 30 for 30 Research Initiative Scheme (project no.  3030\_003) from the Hong Kong University of Science and Technology.
Zhenhui Peng is supported by the Young Scientists Fund of the National Natural Science Foundation of China with Grant No. 62202509.
We are grateful to the anonymous reviewers for their insightful feedback and to our workshop participants for their essential role in facilitating this research.
Last but not least, we appreciate Lennart Nacke's insightful input during the revision process and Cayley MacArthur and Marvin Pafla's insightful review of our work at the University of Waterloo's HCI group meeting.
\end{acks}

\bibliographystyle{ACM-Reference-Format}
\balance
\bibliography{sample-authordraft}


\begin{thebibliography}{100}


\ifx \showCODEN    \undefined \def \showCODEN     #1{\unskip}     \fi
\ifx \showDOI      \undefined \def \showDOI       #1{#1}\fi
\ifx \showISBNx    \undefined \def \showISBNx     #1{\unskip}     \fi
\ifx \showISBNxiii \undefined \def \showISBNxiii  #1{\unskip}     \fi
\ifx \showISSN     \undefined \def \showISSN      #1{\unskip}     \fi
\ifx \showLCCN     \undefined \def \showLCCN      #1{\unskip}     \fi
\ifx \shownote     \undefined \def \shownote      #1{#1}          \fi
\ifx \showarticletitle \undefined \def \showarticletitle #1{#1}   \fi
\ifx \showURL      \undefined \def \showURL       {\relax}        \fi
\providecommand\bibfield[2]{#2}
\providecommand\bibinfo[2]{#2}
\providecommand\natexlab[1]{#1}
\providecommand\showeprint[2][]{arXiv:#2}

\bibitem[Ahn et~al\mbox{.}(2019)]%
        {ahn2019designing}
\bibfield{author}{\bibinfo{person}{June Ahn}, \bibinfo{person}{Fabio Campos},
  \bibinfo{person}{Maria Hays}, {and} \bibinfo{person}{Daniela DiGiacomo}.}
  \bibinfo{year}{2019}\natexlab{}.
\newblock \showarticletitle{Designing in Context: Reaching beyond Usability in
  Learning Analytics Dashboard Design.}
\newblock \bibinfo{journal}{\emph{Journal of Learning Analytics}}
  \bibinfo{volume}{6}, \bibinfo{number}{2} (\bibinfo{year}{2019}),
  \bibinfo{pages}{70--85}.
\newblock


\bibitem[Alvarez et~al\mbox{.}(2020)]%
        {alvarez2020deck}
\bibfield{author}{\bibinfo{person}{Carlos~Prieto Alvarez},
  \bibinfo{person}{Roberto Martinez-Maldonado}, {and} \bibinfo{person}{Simon
  Buckingham~Shum}.} \bibinfo{year}{2020}\natexlab{}.
\newblock \showarticletitle{LA-DECK: A card-based learning analytics co-design
  tool}. In \bibinfo{booktitle}{\emph{Proceedings of the tenth international
  conference on learning analytics \& knowledge}}. \bibinfo{pages}{63--72}.
\newblock


\bibitem[Auger(2010)]%
        {auger2010alternative}
\bibfield{author}{\bibinfo{person}{James Auger}.}
  \bibinfo{year}{2010}\natexlab{}.
\newblock \showarticletitle{Alternative Presents and Speculative Futures:
  Designing fictions through the extrapolation and evasion of product
  lineages.}
\newblock \bibinfo{journal}{\emph{Negotiating futures--Design Fiction.}}
  \bibinfo{volume}{6} (\bibinfo{year}{2010}), \bibinfo{pages}{42--57}.
\newblock


\bibitem[Bach et~al\mbox{.}(2023)]%
        {bach2023challenges}
\bibfield{author}{\bibinfo{person}{Benjamin Bach}, \bibinfo{person}{Mandy
  Keck}, \bibinfo{person}{Fateme Rajabiyazdi}, \bibinfo{person}{Tatiana Losev},
  \bibinfo{person}{Isabel Meirelles}, \bibinfo{person}{Jason Dykes},
  \bibinfo{person}{Robert~S Laramee}, \bibinfo{person}{Mashael AlKadi},
  \bibinfo{person}{Christina Stoiber}, \bibinfo{person}{Samuel Huron},
  {et~al\mbox{.}}} \bibinfo{year}{2023}\natexlab{}.
\newblock \showarticletitle{Challenges and Opportunities in Data Visualization
  Education: A Call to Action}.
\newblock \bibinfo{journal}{\emph{arXiv preprint arXiv:2308.07703}}
  (\bibinfo{year}{2023}).
\newblock


\bibitem[Bae et~al\mbox{.}(2020)]%
        {bae2020spinneret}
\bibfield{author}{\bibinfo{person}{Suyun~Sandra Bae}, \bibinfo{person}{Oh-Hyun
  Kwon}, \bibinfo{person}{Senthil Chandrasegaran}, {and}
  \bibinfo{person}{Kwan-Liu Ma}.} \bibinfo{year}{2020}\natexlab{}.
\newblock \showarticletitle{Spinneret: Aiding creative ideation through
  non-obvious concept associations}. In \bibinfo{booktitle}{\emph{Proceedings
  of the 2020 CHI Conference on Human Factors in Computing Systems}}.
  \bibinfo{pages}{1--13}.
\newblock


\bibitem[Barron et~al\mbox{.}(1998)]%
        {barron1998doing}
\bibfield{author}{\bibinfo{person}{Brigid~JS Barron}, \bibinfo{person}{Daniel~L
  Schwartz}, \bibinfo{person}{Nancy~J Vye}, \bibinfo{person}{Allison Moore},
  \bibinfo{person}{Anthony Petrosino}, \bibinfo{person}{Linda Zech}, {and}
  \bibinfo{person}{John~D Bransford}.} \bibinfo{year}{1998}\natexlab{}.
\newblock \showarticletitle{Doing with understanding: Lessons from research on
  problem-and project-based learning}.
\newblock \bibinfo{journal}{\emph{Journal of the learning sciences}}
  \bibinfo{volume}{7}, \bibinfo{number}{3-4} (\bibinfo{year}{1998}),
  \bibinfo{pages}{271--311}.
\newblock


\bibitem[Bell(2010)]%
        {bell2010project}
\bibfield{author}{\bibinfo{person}{Stephanie Bell}.}
  \bibinfo{year}{2010}\natexlab{}.
\newblock \showarticletitle{Project-based learning for the 21st century: Skills
  for the future}.
\newblock \bibinfo{journal}{\emph{The clearing house}} \bibinfo{volume}{83},
  \bibinfo{number}{2} (\bibinfo{year}{2010}), \bibinfo{pages}{39--43}.
\newblock


\bibitem[Birt et~al\mbox{.}(2016)]%
        {birt2016member}
\bibfield{author}{\bibinfo{person}{Linda Birt}, \bibinfo{person}{Suzanne
  Scott}, \bibinfo{person}{Debbie Cavers}, \bibinfo{person}{Christine
  Campbell}, {and} \bibinfo{person}{Fiona Walter}.}
  \bibinfo{year}{2016}\natexlab{}.
\newblock \showarticletitle{Member checking: a tool to enhance trustworthiness
  or merely a nod to validation?}
\newblock \bibinfo{journal}{\emph{Qualitative health research}}
  \bibinfo{volume}{26}, \bibinfo{number}{13} (\bibinfo{year}{2016}),
  \bibinfo{pages}{1802--1811}.
\newblock


\bibitem[Blumenfeld et~al\mbox{.}(1991)]%
        {blumenfeld1991motivating}
\bibfield{author}{\bibinfo{person}{Phyllis~C Blumenfeld},
  \bibinfo{person}{Elliot Soloway}, \bibinfo{person}{Ronald~W Marx},
  \bibinfo{person}{Joseph~S Krajcik}, \bibinfo{person}{Mark Guzdial}, {and}
  \bibinfo{person}{Annemarie Palincsar}.} \bibinfo{year}{1991}\natexlab{}.
\newblock \showarticletitle{Motivating project-based learning: Sustaining the
  doing, supporting the learning}.
\newblock \bibinfo{journal}{\emph{Educational psychologist}}
  \bibinfo{volume}{26}, \bibinfo{number}{3-4} (\bibinfo{year}{1991}),
  \bibinfo{pages}{369--398}.
\newblock


\bibitem[Braun and Clarke(2006)]%
        {braun2006using}
\bibfield{author}{\bibinfo{person}{Virginia Braun} {and}
  \bibinfo{person}{Victoria Clarke}.} \bibinfo{year}{2006}\natexlab{}.
\newblock \showarticletitle{Using thematic analysis in psychology}.
\newblock \bibinfo{journal}{\emph{Qualitative research in psychology}}
  \bibinfo{volume}{3}, \bibinfo{number}{2} (\bibinfo{year}{2006}),
  \bibinfo{pages}{77--101}.
\newblock


\bibitem[Chan and Hu(2023)]%
        {chan2023students}
\bibfield{author}{\bibinfo{person}{Cecilia Ka~Yuk Chan} {and}
  \bibinfo{person}{Wenjie Hu}.} \bibinfo{year}{2023}\natexlab{}.
\newblock \showarticletitle{Students' Voices on Generative AI: Perceptions,
  Benefits, and Challenges in Higher Education}.
\newblock \bibinfo{journal}{\emph{arXiv preprint arXiv:2305.00290}}
  (\bibinfo{year}{2023}).
\newblock


\bibitem[Chen and Yang(2019)]%
        {chen2019revisiting}
\bibfield{author}{\bibinfo{person}{Cheng-Huan Chen} {and}
  \bibinfo{person}{Yong-Cih Yang}.} \bibinfo{year}{2019}\natexlab{}.
\newblock \showarticletitle{Revisiting the effects of project-based learning on
  students’ academic achievement: A meta-analysis investigating moderators}.
\newblock \bibinfo{journal}{\emph{Educational Research Review}}
  \bibinfo{volume}{26} (\bibinfo{year}{2019}), \bibinfo{pages}{71--81}.
\newblock


\bibitem[Chen et~al\mbox{.}(2023)]%
        {chen2023beyond}
\bibfield{author}{\bibinfo{person}{Zhutian Chen}, \bibinfo{person}{Chenyang
  Zhang}, \bibinfo{person}{Qianwen Wang}, \bibinfo{person}{Jakob Troidl},
  \bibinfo{person}{Simon Warchol}, \bibinfo{person}{Johanna Beyer},
  \bibinfo{person}{Nils Gehlenborg}, {and} \bibinfo{person}{Hanspeter
  Pfister}.} \bibinfo{year}{2023}\natexlab{}.
\newblock \showarticletitle{Beyond Generating Code: Evaluating GPT on a Data
  Visualization Course}.
\newblock \bibinfo{journal}{\emph{arXiv preprint arXiv:2306.02914}}
  (\bibinfo{year}{2023}).
\newblock


\bibitem[Cheon et~al\mbox{.}(2019)]%
        {cheon2019beg}
\bibfield{author}{\bibinfo{person}{EunJeong Cheon}, \bibinfo{person}{Stephen
  Tsung-Han Sher}, \bibinfo{person}{{\v{S}}elma Sabanovi{\'c}}, {and}
  \bibinfo{person}{Norman~Makoto Su}.} \bibinfo{year}{2019}\natexlab{}.
\newblock \showarticletitle{I beg to differ: Soft conflicts in collaborative
  design using design fictions}. In \bibinfo{booktitle}{\emph{Proceedings of
  the 2019 on Designing Interactive Systems Conference}}.
  \bibinfo{pages}{201--214}.
\newblock


\bibitem[Cheon and Su(2018)]%
        {cheon2018futuristic}
\bibfield{author}{\bibinfo{person}{EunJeong Cheon} {and}
  \bibinfo{person}{Norman~Makoto Su}.} \bibinfo{year}{2018}\natexlab{}.
\newblock \showarticletitle{Futuristic autobiographies: Weaving participant
  narratives to elicit values around robots}. In
  \bibinfo{booktitle}{\emph{Proceedings of the 2018 ACM/IEEE International
  Conference on Human-Robot Interaction}}. \bibinfo{pages}{388--397}.
\newblock


\bibitem[Dehouche and Dehouche(2023)]%
        {dehouche2023s}
\bibfield{author}{\bibinfo{person}{Nassim Dehouche} {and}
  \bibinfo{person}{Kullathida Dehouche}.} \bibinfo{year}{2023}\natexlab{}.
\newblock \showarticletitle{What’s in a text-to-image prompt? The potential
  of stable diffusion in visual arts education}.
\newblock \bibinfo{journal}{\emph{Heliyon}} (\bibinfo{year}{2023}).
\newblock


\bibitem[Elsden et~al\mbox{.}(2016)]%
        {elsden2016metadating}
\bibfield{author}{\bibinfo{person}{Chris Elsden}, \bibinfo{person}{Bettina
  Nissen}, \bibinfo{person}{Andrew Garbett}, \bibinfo{person}{David Chatting},
  \bibinfo{person}{David Kirk}, {and} \bibinfo{person}{John Vines}.}
  \bibinfo{year}{2016}\natexlab{}.
\newblock \showarticletitle{Metadating: exploring the romance and future of
  personal data}. In \bibinfo{booktitle}{\emph{Proceedings of the 2016 chi
  conference on human factors in computing systems}}.
  \bibinfo{pages}{685--698}.
\newblock


\bibitem[Fischer(2023)]%
        {fischer2023generative}
\bibfield{author}{\bibinfo{person}{Joel~E Fischer}.}
  \bibinfo{year}{2023}\natexlab{}.
\newblock \showarticletitle{Generative AI Considered Harmful}.
\newblock  (\bibinfo{year}{2023}).
\newblock


\bibitem[Guest et~al\mbox{.}(2011)]%
        {guest2011applied}
\bibfield{author}{\bibinfo{person}{Greg Guest}, \bibinfo{person}{Kathleen~M
  MacQueen}, {and} \bibinfo{person}{Emily~E Namey}.}
  \bibinfo{year}{2011}\natexlab{}.
\newblock \bibinfo{booktitle}{\emph{Applied thematic analysis}}.
\newblock \bibinfo{publisher}{sage publications}.
\newblock


\bibitem[Guo et~al\mbox{.}(2023)]%
        {guo2023close}
\bibfield{author}{\bibinfo{person}{Biyang Guo}, \bibinfo{person}{Xin Zhang},
  \bibinfo{person}{Ziyuan Wang}, \bibinfo{person}{Minqi Jiang},
  \bibinfo{person}{Jinran Nie}, \bibinfo{person}{Yuxuan Ding},
  \bibinfo{person}{Jianwei Yue}, {and} \bibinfo{person}{Yupeng Wu}.}
  \bibinfo{year}{2023}\natexlab{}.
\newblock \bibinfo{title}{How Close is ChatGPT to Human Experts? Comparison
  Corpus, Evaluation, and Detection}.
\newblock
\newblock
\showeprint[arxiv]{2301.07597}~[cs.CL]


\bibitem[Guo et~al\mbox{.}(2020)]%
        {guo2020review}
\bibfield{author}{\bibinfo{person}{Pengyue Guo}, \bibinfo{person}{Nadira Saab},
  \bibinfo{person}{Lysanne~S Post}, {and} \bibinfo{person}{Wilfried Admiraal}.}
  \bibinfo{year}{2020}\natexlab{}.
\newblock \showarticletitle{A review of project-based learning in higher
  education: Student outcomes and measures}.
\newblock \bibinfo{journal}{\emph{International journal of educational
  research}}  \bibinfo{volume}{102} (\bibinfo{year}{2020}),
  \bibinfo{pages}{101586}.
\newblock


\bibitem[{Hadi Mogavi} et~al\mbox{.}(2024)]%
        {mogavi2023exploring}
\bibfield{author}{\bibinfo{person}{Reza {Hadi Mogavi}}, \bibinfo{person}{Chao
  Deng}, \bibinfo{person}{Justin {Juho Kim}}, \bibinfo{person}{Pengyuan Zhou},
  \bibinfo{person}{Young {D. Kwon}}, \bibinfo{person}{Ahmed {Hosny Saleh
  Metwally}}, \bibinfo{person}{Ahmed Tlili}, \bibinfo{person}{Simone
  Bassanelli}, \bibinfo{person}{Antonio Bucchiarone}, \bibinfo{person}{Sujit
  Gujar}, \bibinfo{person}{Lennart~E. Nacke}, {and} \bibinfo{person}{Pan Hui}.}
  \bibinfo{year}{2024}\natexlab{}.
\newblock \showarticletitle{ChatGPT in education: A blessing or a curse? A
  qualitative study exploring early adopters’ utilization and perceptions}.
\newblock \bibinfo{journal}{\emph{Computers in Human Behavior: Artificial
  Humans}} \bibinfo{volume}{2}, \bibinfo{number}{1} (\bibinfo{year}{2024}),
  \bibinfo{pages}{100027}.
\newblock
\showISSN{2949-8821}
\urldef\tempurl%
\url{https://doi.org/10.1016/j.chbah.2023.100027}
\showDOI{\tempurl}


\bibitem[Han et~al\mbox{.}(2022)]%
        {han2022passages}
\bibfield{author}{\bibinfo{person}{Han~L Han}, \bibinfo{person}{Junhang Yu},
  \bibinfo{person}{Raphael Bournet}, \bibinfo{person}{Alexandre Ciorascu},
  \bibinfo{person}{Wendy~E Mackay}, {and} \bibinfo{person}{Michel
  Beaudouin-Lafon}.} \bibinfo{year}{2022}\natexlab{}.
\newblock \showarticletitle{Passages: interacting with text across documents}.
  In \bibinfo{booktitle}{\emph{Proceedings of the 2022 CHI Conference on Human
  Factors in Computing Systems}}. \bibinfo{pages}{1--17}.
\newblock


\bibitem[Hira and Anderson(2021)]%
        {hira2021motivating}
\bibfield{author}{\bibinfo{person}{Avneet Hira} {and} \bibinfo{person}{Emma
  Anderson}.} \bibinfo{year}{2021}\natexlab{}.
\newblock \showarticletitle{Motivating online learning through project-based
  learning during the 2020 COVID-19 pandemic.}
\newblock \bibinfo{journal}{\emph{IAFOR Journal of Education}}
  \bibinfo{volume}{9}, \bibinfo{number}{2} (\bibinfo{year}{2021}),
  \bibinfo{pages}{93--110}.
\newblock


\bibitem[Jakesch et~al\mbox{.}(2019)]%
        {jakesch2019ai}
\bibfield{author}{\bibinfo{person}{Maurice Jakesch}, \bibinfo{person}{Megan
  French}, \bibinfo{person}{Xiao Ma}, \bibinfo{person}{Jeffrey~T Hancock},
  {and} \bibinfo{person}{Mor Naaman}.} \bibinfo{year}{2019}\natexlab{}.
\newblock \showarticletitle{AI-mediated communication: How the perception that
  profile text was written by AI affects trustworthiness}. In
  \bibinfo{booktitle}{\emph{Proceedings of the 2019 CHI Conference on Human
  Factors in Computing Systems}}. \bibinfo{pages}{1--13}.
\newblock


\bibitem[Jonsson and Tholander(2022)]%
        {jonsson2022cracking}
\bibfield{author}{\bibinfo{person}{Martin Jonsson} {and} \bibinfo{person}{Jakob
  Tholander}.} \bibinfo{year}{2022}\natexlab{}.
\newblock \showarticletitle{Cracking the code: Co-coding with AI in creative
  programming education}. In \bibinfo{booktitle}{\emph{Proceedings of the 14th
  Conference on Creativity and Cognition}}. \bibinfo{pages}{5--14}.
\newblock


\bibitem[Kasneci et~al\mbox{.}(2023)]%
        {kasneci2023chatgpt}
\bibfield{author}{\bibinfo{person}{Enkelejda Kasneci}, \bibinfo{person}{Kathrin
  Se{\ss}ler}, \bibinfo{person}{Stefan K{\"u}chemann}, \bibinfo{person}{Maria
  Bannert}, \bibinfo{person}{Daryna Dementieva}, \bibinfo{person}{Frank
  Fischer}, \bibinfo{person}{Urs Gasser}, \bibinfo{person}{Georg Groh},
  \bibinfo{person}{Stephan G{\"u}nnemann}, \bibinfo{person}{Eyke
  H{\"u}llermeier}, {et~al\mbox{.}}} \bibinfo{year}{2023}\natexlab{}.
\newblock \showarticletitle{ChatGPT for good? On opportunities and challenges
  of large language models for education}.
\newblock \bibinfo{journal}{\emph{Learning and individual differences}}
  \bibinfo{volume}{103} (\bibinfo{year}{2023}), \bibinfo{pages}{102274}.
\newblock


\bibitem[Kazemitabaar et~al\mbox{.}(2023)]%
        {kazemitabaar2023studying}
\bibfield{author}{\bibinfo{person}{Majeed Kazemitabaar},
  \bibinfo{person}{Justin Chow}, \bibinfo{person}{Carl Ka~To Ma},
  \bibinfo{person}{Barbara~J Ericson}, \bibinfo{person}{David Weintrop}, {and}
  \bibinfo{person}{Tovi Grossman}.} \bibinfo{year}{2023}\natexlab{}.
\newblock \showarticletitle{Studying the effect of AI Code Generators on
  Supporting Novice Learners in Introductory Programming}. In
  \bibinfo{booktitle}{\emph{Proceedings of the 2023 CHI Conference on Human
  Factors in Computing Systems}}. \bibinfo{pages}{1--23}.
\newblock


\bibitem[Kharrufa et~al\mbox{.}(2017)]%
        {kharrufa2017group}
\bibfield{author}{\bibinfo{person}{Ahmed Kharrufa}, \bibinfo{person}{Sally
  Rix}, \bibinfo{person}{Timur Osadchiy}, \bibinfo{person}{Anne Preston}, {and}
  \bibinfo{person}{Patrick Olivier}.} \bibinfo{year}{2017}\natexlab{}.
\newblock \showarticletitle{Group Spinner: recognizing and visualizing learning
  in the classroom for reflection, communication, and planning}. In
  \bibinfo{booktitle}{\emph{Proceedings of the 2017 CHI Conference on Human
  Factors in Computing Systems}}. \bibinfo{pages}{5556--5567}.
\newblock


\bibitem[Kim et~al\mbox{.}(2023)]%
        {kim2023one}
\bibfield{author}{\bibinfo{person}{Taenyun Kim}, \bibinfo{person}{Maria~D
  Molina}, \bibinfo{person}{Minjin Rheu}, \bibinfo{person}{Emily~S Zhan}, {and}
  \bibinfo{person}{Wei Peng}.} \bibinfo{year}{2023}\natexlab{}.
\newblock \showarticletitle{One AI Does Not Fit All: A Cluster Analysis of the
  Laypeople’s Perception of AI Roles}. In
  \bibinfo{booktitle}{\emph{Proceedings of the 2023 CHI Conference on Human
  Factors in Computing Systems}}. \bibinfo{pages}{1--20}.
\newblock


\bibitem[Kokotsaki et~al\mbox{.}(2016)]%
        {kokotsaki2016project}
\bibfield{author}{\bibinfo{person}{Dimitra Kokotsaki},
  \bibinfo{person}{Victoria Menzies}, {and} \bibinfo{person}{Andy Wiggins}.}
  \bibinfo{year}{2016}\natexlab{}.
\newblock \showarticletitle{Project-based learning: A review of the
  literature}.
\newblock \bibinfo{journal}{\emph{Improving schools}} \bibinfo{volume}{19},
  \bibinfo{number}{3} (\bibinfo{year}{2016}), \bibinfo{pages}{267--277}.
\newblock


\bibitem[Krajcik and Blumenfeld(2006)]%
        {krajcik2006project}
\bibfield{author}{\bibinfo{person}{Joseph~S Krajcik} {and}
  \bibinfo{person}{Phyllis~C Blumenfeld}.} \bibinfo{year}{2006}\natexlab{}.
\newblock \bibinfo{booktitle}{\emph{Project-based learning}}.
\newblock \bibinfo{publisher}{na}.
\newblock


\bibitem[Kung et~al\mbox{.}(2023)]%
        {kung2023performance}
\bibfield{author}{\bibinfo{person}{Tiffany~H Kung}, \bibinfo{person}{Morgan
  Cheatham}, \bibinfo{person}{Arielle Medenilla}, \bibinfo{person}{Czarina
  Sillos}, \bibinfo{person}{Lorie De~Leon}, \bibinfo{person}{Camille
  Elepa{\~n}o}, \bibinfo{person}{Maria Madriaga}, \bibinfo{person}{Rimel
  Aggabao}, \bibinfo{person}{Giezel Diaz-Candido}, \bibinfo{person}{James
  Maningo}, {et~al\mbox{.}}} \bibinfo{year}{2023}\natexlab{}.
\newblock \showarticletitle{Performance of ChatGPT on USMLE: Potential for
  AI-assisted medical education using large language models}.
\newblock \bibinfo{journal}{\emph{PLoS digital health}} \bibinfo{volume}{2},
  \bibinfo{number}{2} (\bibinfo{year}{2023}), \bibinfo{pages}{e0000198}.
\newblock


\bibitem[Lai et~al\mbox{.}(2022)]%
        {lai2022human}
\bibfield{author}{\bibinfo{person}{Vivian Lai}, \bibinfo{person}{Samuel
  Carton}, \bibinfo{person}{Rajat Bhatnagar}, \bibinfo{person}{Q~Vera Liao},
  \bibinfo{person}{Yunfeng Zhang}, {and} \bibinfo{person}{Chenhao Tan}.}
  \bibinfo{year}{2022}\natexlab{}.
\newblock \showarticletitle{Human-ai collaboration via conditional delegation:
  A case study of content moderation}. In \bibinfo{booktitle}{\emph{Proceedings
  of the 2022 CHI Conference on Human Factors in Computing Systems}}.
  \bibinfo{pages}{1--18}.
\newblock


\bibitem[Lamb(2003)]%
        {lamb2003project}
\bibfield{author}{\bibinfo{person}{Darren~Hayes Lamb}.}
  \bibinfo{year}{2003}\natexlab{}.
\newblock \showarticletitle{Project based learning in an applied construction
  curriculum}.
\newblock  (\bibinfo{year}{2003}).
\newblock


\bibitem[Lau and Guo(2023)]%
        {lau2023ban}
\bibfield{author}{\bibinfo{person}{Sam Lau} {and} \bibinfo{person}{Philip~J
  Guo}.} \bibinfo{year}{2023}\natexlab{}.
\newblock \showarticletitle{From" Ban It Till We Understand It" to" Resistance
  is Futile": How University Programming Instructors Plan to Adapt as More
  Students Use AI Code Generation and Explanation Tools such as ChatGPT and
  GitHub Copilot}.
\newblock  (\bibinfo{year}{2023}).
\newblock


\bibitem[Lee et~al\mbox{.}(2022)]%
        {lee2022impacts}
\bibfield{author}{\bibinfo{person}{Yen-Fen Lee}, \bibinfo{person}{Gwo-Jen
  Hwang}, {and} \bibinfo{person}{Pei-Ying Chen}.}
  \bibinfo{year}{2022}\natexlab{}.
\newblock \showarticletitle{Impacts of an AI-based cha bot on college
  students’ after-class review, academic performance, self-efficacy, learning
  attitude, and motivation}.
\newblock \bibinfo{journal}{\emph{Educational technology research and
  development}} \bibinfo{volume}{70}, \bibinfo{number}{5}
  (\bibinfo{year}{2022}), \bibinfo{pages}{1843--1865}.
\newblock


\bibitem[Li et~al\mbox{.}(2011)]%
        {li2011personal}
\bibfield{author}{\bibinfo{person}{Ian Li}, \bibinfo{person}{Anind Dey},
  \bibinfo{person}{Jodi Forlizzi}, \bibinfo{person}{Kristina H{\"o}{\"o}k},
  {and} \bibinfo{person}{Yevgeniy Medynskiy}.} \bibinfo{year}{2011}\natexlab{}.
\newblock \showarticletitle{Personal informatics and HCI: design, theory, and
  social implications}.
\newblock In \bibinfo{booktitle}{\emph{CHI'11 Extended Abstracts on Human
  Factors in Computing Systems}}. \bibinfo{pages}{2417--2420}.
\newblock


\bibitem[Licklider(1960)]%
        {licklider1960man}
\bibfield{author}{\bibinfo{person}{Joseph~CR Licklider}.}
  \bibinfo{year}{1960}\natexlab{}.
\newblock \showarticletitle{Man-computer symbiosis}.
\newblock \bibinfo{journal}{\emph{IRE transactions on human factors in
  electronics}} \bibinfo{number}{1} (\bibinfo{year}{1960}),
  \bibinfo{pages}{4--11}.
\newblock


\bibitem[Lindley and Coulton(2015)]%
        {lindley2015back}
\bibfield{author}{\bibinfo{person}{Joseph Lindley} {and} \bibinfo{person}{Paul
  Coulton}.} \bibinfo{year}{2015}\natexlab{}.
\newblock \showarticletitle{Back to the future: 10 years of design fiction}. In
  \bibinfo{booktitle}{\emph{Proceedings of the 2015 British HCI conference}}.
  \bibinfo{pages}{210--211}.
\newblock


\bibitem[Lindley et~al\mbox{.}(2018)]%
        {lindley2018exploring}
\bibfield{author}{\bibinfo{person}{Si{\^a}n~E Lindley}, \bibinfo{person}{Gavin
  Smyth}, \bibinfo{person}{Robert Corish}, \bibinfo{person}{Anastasia
  Loukianov}, \bibinfo{person}{Michael Golembewski}, \bibinfo{person}{Ewa~A
  Luger}, {and} \bibinfo{person}{Abigail Sellen}.}
  \bibinfo{year}{2018}\natexlab{}.
\newblock \showarticletitle{Exploring new metaphors for a networked world
  through the file biography}. In \bibinfo{booktitle}{\emph{Proceedings of the
  2018 CHI Conference on Human Factors in Computing Systems}}.
  \bibinfo{pages}{1--12}.
\newblock


\bibitem[Long and Magerko(2020)]%
        {long2020ai}
\bibfield{author}{\bibinfo{person}{Duri Long} {and} \bibinfo{person}{Brian
  Magerko}.} \bibinfo{year}{2020}\natexlab{}.
\newblock \showarticletitle{What is AI literacy? Competencies and design
  considerations}. In \bibinfo{booktitle}{\emph{Proceedings of the 2020 CHI
  conference on human factors in computing systems}}. \bibinfo{pages}{1--16}.
\newblock


\bibitem[Luria(2023)]%
        {luria2023co}
\bibfield{author}{\bibinfo{person}{Michal Luria}.}
  \bibinfo{year}{2023}\natexlab{}.
\newblock \showarticletitle{Co-Design Perspectives on Algorithm Transparency
  Reporting: Guidelines and Prototypes}. In
  \bibinfo{booktitle}{\emph{Proceedings of the 2023 ACM Conference on Fairness,
  Accountability, and Transparency}}. \bibinfo{pages}{1076--1087}.
\newblock


\bibitem[Ma et~al\mbox{.}(2022)]%
        {ma2022glancee}
\bibfield{author}{\bibinfo{person}{Shuai Ma}, \bibinfo{person}{Taichang Zhou},
  \bibinfo{person}{Fei Nie}, {and} \bibinfo{person}{Xiaojuan Ma}.}
  \bibinfo{year}{2022}\natexlab{}.
\newblock \showarticletitle{Glancee: An adaptable system for instructors to
  grasp student learning status in synchronous online classes}. In
  \bibinfo{booktitle}{\emph{Proceedings of the 2022 CHI Conference on Human
  Factors in Computing Systems}}. \bibinfo{pages}{1--25}.
\newblock


\bibitem[Mhlanga(2023)]%
        {mhlanga2023open}
\bibfield{author}{\bibinfo{person}{David Mhlanga}.}
  \bibinfo{year}{2023}\natexlab{}.
\newblock \showarticletitle{Open AI in education, the responsible and ethical
  use of ChatGPT towards lifelong learning}.
\newblock \bibinfo{journal}{\emph{Education, the Responsible and Ethical Use of
  ChatGPT Towards Lifelong Learning (February 11, 2023)}}
  (\bibinfo{year}{2023}).
\newblock


\bibitem[Midjourney(2023)]%
        {midjourney}
\bibfield{author}{\bibinfo{person}{Midjourney}.}
  \bibinfo{year}{2023}\natexlab{}.
\newblock \bibinfo{title}{Midjourney}.
\newblock
\newblock
\urldef\tempurl%
\url{https://www.midjourney.com/}
\showURL{%
\tempurl}
\newblock
\shownote{https://www.midjourney.com/}.


\bibitem[Mogavi et~al\mbox{.}(2021)]%
        {mogavi2021characterizing}
\bibfield{author}{\bibinfo{person}{Reza~Hadi Mogavi}, \bibinfo{person}{Xiaojuan
  Ma}, {and} \bibinfo{person}{Pan Hui}.} \bibinfo{year}{2021}\natexlab{}.
\newblock \showarticletitle{Characterizing student engagement moods for dropout
  prediction in question pool websites}.
\newblock \bibinfo{journal}{\emph{arXiv preprint arXiv:2102.00423}}
  (\bibinfo{year}{2021}).
\newblock


\bibitem[Mollick and Mollick(2023)]%
        {mollick2023assigning}
\bibfield{author}{\bibinfo{person}{Ethan Mollick} {and} \bibinfo{person}{Lilach
  Mollick}.} \bibinfo{year}{2023}\natexlab{}.
\newblock \showarticletitle{Assigning AI: Seven Approaches for Students, with
  Prompts}.
\newblock \bibinfo{journal}{\emph{arXiv preprint arXiv:2306.10052}}
  (\bibinfo{year}{2023}).
\newblock


\bibitem[Murgia et~al\mbox{.}(2023)]%
        {murgia2023chatgpt}
\bibfield{author}{\bibinfo{person}{Emiliana Murgia}, \bibinfo{person}{Zahra
  Abbasiantaeb}, \bibinfo{person}{Mohammad Aliannejadi}, \bibinfo{person}{Theo
  Huibers}, \bibinfo{person}{Monica Landoni}, {and}
  \bibinfo{person}{Maria~Soledad Pera}.} \bibinfo{year}{2023}\natexlab{}.
\newblock \showarticletitle{ChatGPT in the Classroom: A Preliminary Exploration
  on the Feasibility of Adapting ChatGPT to Support Children’s Information
  Discovery}. In \bibinfo{booktitle}{\emph{Adjunct Proceedings of the 31st ACM
  Conference on User Modeling, Adaptation and Personalization}}.
  \bibinfo{pages}{22--27}.
\newblock


\bibitem[Neumann et~al\mbox{.}(2023)]%
        {neumann2023we}
\bibfield{author}{\bibinfo{person}{Michael Neumann}, \bibinfo{person}{Maria
  Rauschenberger}, {and} \bibinfo{person}{Eva-Maria Sch{\"o}n}.}
  \bibinfo{year}{2023}\natexlab{}.
\newblock \showarticletitle{“We Need To Talk About ChatGPT”: The Future of
  AI and Higher Education}.
\newblock  (\bibinfo{year}{2023}).
\newblock


\bibitem[of~Cambridge(2023)]%
        {cambridge2023}
\bibfield{author}{\bibinfo{person}{University of Cambridge}.}
  \bibinfo{year}{2023}\natexlab{}.
\newblock
\newblock
\urldef\tempurl%
\url{https://www.plagiarism.admin.cam.ac.uk/what-academic-misconduct/artificial-intelligence}
\showURL{%
Retrieved September 11, 2023 from \tempurl}


\bibitem[of~Hong~Kong(2023)]%
        {hku2023}
\bibfield{author}{\bibinfo{person}{The~University of Hong~Kong}.}
  \bibinfo{year}{2023}\natexlab{}.
\newblock
\newblock
\urldef\tempurl%
\url{https://tl.hku.hk/2023/02/about-chatgpt/}
\showURL{%
Retrieved September 11, 2023 from \tempurl}


\bibitem[of~Oxford(2023)]%
        {oxford2023}
\bibfield{author}{\bibinfo{person}{University of Oxford}.}
  \bibinfo{year}{2023}\natexlab{}.
\newblock \bibinfo{title}{Unauthorised use of AI in exams and assessment}.
\newblock
\newblock
\urldef\tempurl%
\url{https://academic.admin.ox.ac.uk/article/unauthorised-use-of-ai-in-exams-and-assessment}
\showURL{%
Retrieved September 11, 2023 from \tempurl}


\bibitem[of~Science and Technology(2023)]%
        {hkust2023}
\bibfield{author}{\bibinfo{person}{The Hong Kong~University of Science} {and}
  \bibinfo{person}{Technology}.} \bibinfo{year}{2023}\natexlab{}.
\newblock
\newblock
\urldef\tempurl%
\url{https://cei.hkust.edu.hk/en-hk/education-innovation/generative-ai-education/guidelines-and-policies}
\showURL{%
Retrieved September 11, 2023 from \tempurl}


\bibitem[of~Washington(2023)]%
        {uw2023}
\bibfield{author}{\bibinfo{person}{University of Washington}.}
  \bibinfo{year}{2023}\natexlab{}.
\newblock \bibinfo{title}{ChatGPT and other AI-based tools}.
\newblock
\newblock
\urldef\tempurl%
\url{https://teaching.washington.edu/course-design/chatgpt/}
\showURL{%
Retrieved September 11, 2023 from \tempurl}


\bibitem[Olson and Kellogg(2014)]%
        {olson2014ways}
\bibfield{author}{\bibinfo{person}{Judith~S Olson} {and}
  \bibinfo{person}{Wendy~A Kellogg}.} \bibinfo{year}{2014}\natexlab{}.
\newblock \bibinfo{booktitle}{\emph{Ways of Knowing in HCI}}.
  Vol.~\bibinfo{volume}{2}.
\newblock \bibinfo{publisher}{Springer}.
\newblock


\bibitem[OpenAI(2023a)]%
        {ChatGPT2023}
\bibfield{author}{\bibinfo{person}{OpenAI}.} \bibinfo{year}{2023}\natexlab{a}.
\newblock \bibinfo{title}{ChatGPT}.
\newblock
\newblock
\urldef\tempurl%
\url{https://openai.com/chatgpt}
\showURL{%
\tempurl}
\newblock
\shownote{https://openai.com/chatgpt}.


\bibitem[OpenAI(2023b)]%
        {openai2023edufqa}
\bibfield{author}{\bibinfo{person}{OpenAI}.} \bibinfo{year}{2023}\natexlab{b}.
\newblock \bibinfo{title}{Educator FAQ}.
\newblock
\newblock
\urldef\tempurl%
\url{https://help.openai.com/en/collections/5929286-educator-faq}
\showURL{%
Retrieved September 11, 2023 from \tempurl}
\newblock
\shownote{https://help.openai.com/en/collections/5929286-educator-faq}.


\bibitem[Pardos et~al\mbox{.}(2023)]%
        {pardos2023oatutor}
\bibfield{author}{\bibinfo{person}{Zachary~A Pardos}, \bibinfo{person}{Matthew
  Tang}, \bibinfo{person}{Ioannis Anastasopoulos}, \bibinfo{person}{Shreya~K
  Sheel}, {and} \bibinfo{person}{Ethan Zhang}.}
  \bibinfo{year}{2023}\natexlab{}.
\newblock \showarticletitle{OATutor: An Open-source Adaptive Tutoring System
  and Curated Content Library for Learning Sciences Research}. In
  \bibinfo{booktitle}{\emph{Proceedings of the 2023 CHI Conference on Human
  Factors in Computing Systems}}. \bibinfo{pages}{1--17}.
\newblock


\bibitem[Peng et~al\mbox{.}(2023)]%
        {peng2023storyfier}
\bibfield{author}{\bibinfo{person}{Zhenhui Peng}, \bibinfo{person}{Xingbo
  Wang}, \bibinfo{person}{Qiushi Han}, \bibinfo{person}{Junkai Zhu},
  \bibinfo{person}{Xiaojuan Ma}, {and} \bibinfo{person}{Huamin Qu}.}
  \bibinfo{year}{2023}\natexlab{}.
\newblock \showarticletitle{Storyfier: Exploring Vocabulary Learning Support
  with Text Generation Models}. In \bibinfo{booktitle}{\emph{Proceedings of the
  36th Annual ACM Symposium on User Interface Software and Technology}}.
  \bibinfo{pages}{1--16}.
\newblock


\bibitem[P{\'e}rez and Rubio(2020)]%
        {perez2020project}
\bibfield{author}{\bibinfo{person}{Beatriz P{\'e}rez} {and}
  \bibinfo{person}{{\'A}ngel~L Rubio}.} \bibinfo{year}{2020}\natexlab{}.
\newblock \showarticletitle{A project-based learning approach for enhancing
  learning skills and motivation in software engineering}. In
  \bibinfo{booktitle}{\emph{Proceedings of the 51st ACM technical symposium on
  computer science education}}. \bibinfo{pages}{309--315}.
\newblock


\bibitem[Piccolo et~al\mbox{.}(2023)]%
        {piccolo2023interaction}
\bibfield{author}{\bibinfo{person}{Lara Piccolo}, \bibinfo{person}{Daniel
  Buzzo}, \bibinfo{person}{Martin Knobel}, \bibinfo{person}{Prasanna
  Gunasekera}, {and} \bibinfo{person}{Tina Papathoma}.}
  \bibinfo{year}{2023}\natexlab{}.
\newblock \showarticletitle{Interaction Design as Project-Based Learning:
  Perspectives for Unsolved Challenges}. In
  \bibinfo{booktitle}{\emph{Proceedings of the 5th Annual Symposium on HCI
  Education}}. \bibinfo{pages}{59--67}.
\newblock


\bibitem[Prieto-Alvarez et~al\mbox{.}(2018a)]%
        {prieto2018co}
\bibfield{author}{\bibinfo{person}{Carlos~G Prieto-Alvarez},
  \bibinfo{person}{Roberto Martinez-Maldonado}, {and}
  \bibinfo{person}{Theresa~Dirndorfer Anderson}.}
  \bibinfo{year}{2018}\natexlab{a}.
\newblock \showarticletitle{Co-designing learning analytics tools with
  learners}.
\newblock In \bibinfo{booktitle}{\emph{Learning analytics in the classroom}}.
  \bibinfo{publisher}{Routledge}, \bibinfo{pages}{93--110}.
\newblock


\bibitem[Prieto-Alvarez et~al\mbox{.}(2018b)]%
        {prieto2018mapping}
\bibfield{author}{\bibinfo{person}{Carlos~Gerardo Prieto-Alvarez},
  \bibinfo{person}{Roberto Martinez-Maldonado}, {and} \bibinfo{person}{Simon
  Buckingham~Shum}.} \bibinfo{year}{2018}\natexlab{b}.
\newblock \showarticletitle{Mapping learner-data journeys: Evolution of a
  visual co-design tool}. In \bibinfo{booktitle}{\emph{Proceedings of the 30th
  Australian conference on computer-human interaction}}.
  \bibinfo{pages}{205--214}.
\newblock


\bibitem[Qadir(2023)]%
        {qadir2023engineering}
\bibfield{author}{\bibinfo{person}{Junaid Qadir}.}
  \bibinfo{year}{2023}\natexlab{}.
\newblock \showarticletitle{Engineering education in the era of ChatGPT:
  Promise and pitfalls of generative AI for education}. In
  \bibinfo{booktitle}{\emph{2023 IEEE Global Engineering Education Conference
  (EDUCON)}}. IEEE, \bibinfo{pages}{1--9}.
\newblock


\bibitem[Rasul et~al\mbox{.}(2023)]%
        {rasul2023role}
\bibfield{author}{\bibinfo{person}{Tareq Rasul}, \bibinfo{person}{Sumesh Nair},
  \bibinfo{person}{Diane Kalendra}, \bibinfo{person}{Mulyadi Robin},
  \bibinfo{person}{Fernando de Oliveira~Santini},
  \bibinfo{person}{Wagner~Junior Ladeira}, \bibinfo{person}{Mingwei Sun},
  \bibinfo{person}{Ingrid Day}, \bibinfo{person}{Raouf~Ahmad Rather}, {and}
  \bibinfo{person}{Liz Heathcote}.} \bibinfo{year}{2023}\natexlab{}.
\newblock \showarticletitle{The role of ChatGPT in higher education: Benefits,
  challenges, and future research directions}.
\newblock \bibinfo{journal}{\emph{Journal of Applied Learning and Teaching}}
  \bibinfo{volume}{6}, \bibinfo{number}{1} (\bibinfo{year}{2023}).
\newblock


\bibitem[Richardson and Kharrufa(2020)]%
        {richardson2020we}
\bibfield{author}{\bibinfo{person}{Dan Richardson} {and} \bibinfo{person}{Ahmed
  Kharrufa}.} \bibinfo{year}{2020}\natexlab{}.
\newblock \showarticletitle{We are the greatest showmen: Configuring a
  framework for project-based mobile learning}. In
  \bibinfo{booktitle}{\emph{Proceedings of the 2020 CHI Conference on Human
  Factors in Computing Systems}}. \bibinfo{pages}{1--12}.
\newblock


\bibitem[Romrell et~al\mbox{.}(2014)]%
        {romrell2014samr}
\bibfield{author}{\bibinfo{person}{Danae Romrell}, \bibinfo{person}{Lisa
  Kidder}, {and} \bibinfo{person}{Emma Wood}.} \bibinfo{year}{2014}\natexlab{}.
\newblock \showarticletitle{The SAMR model as a framework for evaluating
  mLearning}.
\newblock \bibinfo{journal}{\emph{Online Learning Journal}}
  \bibinfo{volume}{18}, \bibinfo{number}{2} (\bibinfo{year}{2014}).
\newblock


\bibitem[Rong et~al\mbox{.}(2023)]%
        {rong2023understanding}
\bibfield{author}{\bibinfo{person}{Ethan~Z Rong}, \bibinfo{person}{Mo~Morgana
  Zhou}, \bibinfo{person}{Ge Gao}, {and} \bibinfo{person}{Zhicong Lu}.}
  \bibinfo{year}{2023}\natexlab{}.
\newblock \showarticletitle{Understanding Personal Data Tracking and
  Sensemaking Practices for Self-Directed Learning in Non-classroom and
  Non-computer-based Contexts}. In \bibinfo{booktitle}{\emph{Proceedings of the
  2023 CHI Conference on Human Factors in Computing Systems}}.
  \bibinfo{pages}{1--16}.
\newblock


\bibitem[Rudolph et~al\mbox{.}(2023)]%
        {rudolph2023chatgpt}
\bibfield{author}{\bibinfo{person}{J{\"u}rgen Rudolph}, \bibinfo{person}{Samson
  Tan}, {and} \bibinfo{person}{Shannon Tan}.} \bibinfo{year}{2023}\natexlab{}.
\newblock \showarticletitle{ChatGPT: Bullshit spewer or the end of traditional
  assessments in higher education?}
\newblock \bibinfo{journal}{\emph{Journal of Applied Learning and Teaching}}
  \bibinfo{volume}{6}, \bibinfo{number}{1} (\bibinfo{year}{2023}).
\newblock


\bibitem[Sarmiento and Wise(2022)]%
        {sarmiento2022participatory}
\bibfield{author}{\bibinfo{person}{Juan~Pablo Sarmiento} {and}
  \bibinfo{person}{Alyssa~Friend Wise}.} \bibinfo{year}{2022}\natexlab{}.
\newblock \showarticletitle{Participatory and co-design of learning analytics:
  An initial review of the literature}. In \bibinfo{booktitle}{\emph{LAK22:
  12th international learning analytics and knowledge conference}}.
  \bibinfo{pages}{535--541}.
\newblock


\bibitem[Schuler and Namioka(1993)]%
        {schuler1993participatory}
\bibfield{author}{\bibinfo{person}{Douglas Schuler} {and} \bibinfo{person}{Aki
  Namioka}.} \bibinfo{year}{1993}\natexlab{}.
\newblock \bibinfo{booktitle}{\emph{Participatory design: Principles and
  practices}}.
\newblock \bibinfo{publisher}{CRC Press}.
\newblock


\bibitem[Shi et~al\mbox{.}(2023)]%
        {shi2023retrolens}
\bibfield{author}{\bibinfo{person}{Chuhan Shi}, \bibinfo{person}{Yicheng Hu},
  \bibinfo{person}{Shenan Wang}, \bibinfo{person}{Shuai Ma},
  \bibinfo{person}{Chengbo Zheng}, \bibinfo{person}{Xiaojuan Ma}, {and}
  \bibinfo{person}{Qiong Luo}.} \bibinfo{year}{2023}\natexlab{}.
\newblock \showarticletitle{RetroLens: A Human-AI Collaborative System for
  Multi-step Retrosynthetic Route Planning}. In
  \bibinfo{booktitle}{\emph{Proceedings of the 2023 CHI Conference on Human
  Factors in Computing Systems}}. \bibinfo{pages}{1--20}.
\newblock


\bibitem[Smith(2018)]%
        {smith2018generalizability}
\bibfield{author}{\bibinfo{person}{Brett Smith}.}
  \bibinfo{year}{2018}\natexlab{}.
\newblock \showarticletitle{Generalizability in qualitative research:
  Misunderstandings, opportunities and recommendations for the sport and
  exercise sciences}.
\newblock \bibinfo{journal}{\emph{Qualitative research in sport, exercise and
  health}} \bibinfo{volume}{10}, \bibinfo{number}{1} (\bibinfo{year}{2018}),
  \bibinfo{pages}{137--149}.
\newblock


\bibitem[Sterman et~al\mbox{.}(2023)]%
        {sterman2023kaleidoscope}
\bibfield{author}{\bibinfo{person}{Sarah Sterman}, \bibinfo{person}{Molly~Jane
  Nicholas}, \bibinfo{person}{Janaki Vivrekar}, \bibinfo{person}{Jessica~R
  Mindel}, {and} \bibinfo{person}{Eric Paulos}.}
  \bibinfo{year}{2023}\natexlab{}.
\newblock \showarticletitle{Kaleidoscope: A Reflective Documentation Tool for a
  User Interface Design Course}. In \bibinfo{booktitle}{\emph{Proceedings of
  the 2023 CHI Conference on Human Factors in Computing Systems}}.
  \bibinfo{pages}{1--19}.
\newblock


\bibitem[Thomas(2000)]%
        {thomas2000review}
\bibfield{author}{\bibinfo{person}{John Thomas}.}
  \bibinfo{year}{2000}\natexlab{}.
\newblock \showarticletitle{A Review of Research on Project-Based Learning}.
\newblock  (\bibinfo{date}{01} \bibinfo{year}{2000}).
\newblock


\bibitem[Tian et~al\mbox{.}(2023)]%
        {tian2023last}
\bibfield{author}{\bibinfo{person}{Yao Tian}, \bibinfo{person}{Chengwei Tong},
  \bibinfo{person}{Lik-Hang Lee}, \bibinfo{person}{Reza~Hadi Mogavi},
  \bibinfo{person}{Yong Liao}, {and} \bibinfo{person}{Pengyuan Zhou}.}
  \bibinfo{year}{2023}\natexlab{}.
\newblock \showarticletitle{Last Week with ChatGPT: A Weibo Study on Social
  Perspective regarding ChatGPT for Education and Beyond}.
\newblock \bibinfo{journal}{\emph{arXiv preprint arXiv:2306.04325}}
  (\bibinfo{year}{2023}).
\newblock


\bibitem[Varanasi and Goyal(2023)]%
        {varanasi2023currently}
\bibfield{author}{\bibinfo{person}{Rama~Adithya Varanasi} {and}
  \bibinfo{person}{Nitesh Goyal}.} \bibinfo{year}{2023}\natexlab{}.
\newblock \showarticletitle{“It is currently hodgepodge”: Examining AI/ML
  Practitioners’ Challenges during Co-production of Responsible AI Values}.
  In \bibinfo{booktitle}{\emph{Proceedings of the 2023 CHI Conference on Human
  Factors in Computing Systems}}. \bibinfo{pages}{1--17}.
\newblock


\bibitem[Wang et~al\mbox{.}(2019)]%
        {wang2019human}
\bibfield{author}{\bibinfo{person}{Dakuo Wang}, \bibinfo{person}{Justin~D
  Weisz}, \bibinfo{person}{Michael Muller}, \bibinfo{person}{Parikshit Ram},
  \bibinfo{person}{Werner Geyer}, \bibinfo{person}{Casey Dugan},
  \bibinfo{person}{Yla Tausczik}, \bibinfo{person}{Horst Samulowitz}, {and}
  \bibinfo{person}{Alexander Gray}.} \bibinfo{year}{2019}\natexlab{}.
\newblock \showarticletitle{Human-ai collaboration in data science: Exploring
  data scientists' perceptions of automated ai}.
\newblock \bibinfo{journal}{\emph{Proceedings of the ACM on human-computer
  interaction}} \bibinfo{volume}{3}, \bibinfo{number}{CSCW}
  (\bibinfo{year}{2019}), \bibinfo{pages}{1--24}.
\newblock


\bibitem[Wang et~al\mbox{.}(2022)]%
        {wang2022co}
\bibfield{author}{\bibinfo{person}{Qiaosi Wang}, \bibinfo{person}{Shan Jing},
  {and} \bibinfo{person}{Ashok~K Goel}.} \bibinfo{year}{2022}\natexlab{}.
\newblock \showarticletitle{Co-Designing AI Agents to Support Social
  Connectedness Among Online Learners: Functionalities, Social Characteristics,
  and Ethical Challenges}. In \bibinfo{booktitle}{\emph{Designing Interactive
  Systems Conference}}. \bibinfo{pages}{541--556}.
\newblock


\bibitem[Wang et~al\mbox{.}(2023)]%
        {wang2023designing}
\bibfield{author}{\bibinfo{person}{Qiaosi Wang}, \bibinfo{person}{Michael
  Madaio}, \bibinfo{person}{Shaun Kane}, \bibinfo{person}{Shivani Kapania},
  \bibinfo{person}{Michael Terry}, {and} \bibinfo{person}{Lauren Wilcox}.}
  \bibinfo{year}{2023}\natexlab{}.
\newblock \showarticletitle{Designing Responsible AI: Adaptations of UX
  Practice to Meet Responsible AI Challenges}. In
  \bibinfo{booktitle}{\emph{Proceedings of the 2023 CHI Conference on Human
  Factors in Computing Systems}}. \bibinfo{pages}{1--16}.
\newblock


\bibitem[Wang et~al\mbox{.}(2021)]%
        {wang2021towards}
\bibfield{author}{\bibinfo{person}{Qiaosi Wang}, \bibinfo{person}{Koustuv
  Saha}, \bibinfo{person}{Eric Gregori}, \bibinfo{person}{David Joyner}, {and}
  \bibinfo{person}{Ashok Goel}.} \bibinfo{year}{2021}\natexlab{}.
\newblock \showarticletitle{Towards mutual theory of mind in human-ai
  interaction: How language reflects what students perceive about a virtual
  teaching assistant}. In \bibinfo{booktitle}{\emph{Proceedings of the 2021 CHI
  conference on human factors in computing systems}}. \bibinfo{pages}{1--14}.
\newblock


\bibitem[Wei et~al\mbox{.}(2022a)]%
        {wei2022emergent}
\bibfield{author}{\bibinfo{person}{Jason Wei}, \bibinfo{person}{Yi Tay},
  \bibinfo{person}{Rishi Bommasani}, \bibinfo{person}{Colin Raffel},
  \bibinfo{person}{Barret Zoph}, \bibinfo{person}{Sebastian Borgeaud},
  \bibinfo{person}{Dani Yogatama}, \bibinfo{person}{Maarten Bosma},
  \bibinfo{person}{Denny Zhou}, \bibinfo{person}{Donald Metzler},
  {et~al\mbox{.}}} \bibinfo{year}{2022}\natexlab{a}.
\newblock \showarticletitle{Emergent abilities of large language models}.
\newblock \bibinfo{journal}{\emph{arXiv preprint arXiv:2206.07682}}
  (\bibinfo{year}{2022}).
\newblock


\bibitem[Wei et~al\mbox{.}(2022b)]%
        {wei2022chain}
\bibfield{author}{\bibinfo{person}{Jason Wei}, \bibinfo{person}{Xuezhi Wang},
  \bibinfo{person}{Dale Schuurmans}, \bibinfo{person}{Maarten Bosma},
  \bibinfo{person}{Fei Xia}, \bibinfo{person}{Ed Chi}, \bibinfo{person}{Quoc~V
  Le}, \bibinfo{person}{Denny Zhou}, {et~al\mbox{.}}}
  \bibinfo{year}{2022}\natexlab{b}.
\newblock \showarticletitle{Chain-of-thought prompting elicits reasoning in
  large language models}.
\newblock \bibinfo{journal}{\emph{Advances in Neural Information Processing
  Systems}}  \bibinfo{volume}{35} (\bibinfo{year}{2022}),
  \bibinfo{pages}{24824--24837}.
\newblock


\bibitem[Weitekamp et~al\mbox{.}(2020)]%
        {weitekamp2020interaction}
\bibfield{author}{\bibinfo{person}{Daniel Weitekamp}, \bibinfo{person}{Erik
  Harpstead}, {and} \bibinfo{person}{Ken~R Koedinger}.}
  \bibinfo{year}{2020}\natexlab{}.
\newblock \showarticletitle{An interaction design for machine teaching to
  develop AI tutors}. In \bibinfo{booktitle}{\emph{Proceedings of the 2020 CHI
  conference on human factors in computing systems}}. \bibinfo{pages}{1--11}.
\newblock


\bibitem[Wirfs-Brock et~al\mbox{.}(2020)]%
        {wirfs2020giving}
\bibfield{author}{\bibinfo{person}{Jordan Wirfs-Brock}, \bibinfo{person}{Sarah
  Mennicken}, {and} \bibinfo{person}{Jennifer Thom}.}
  \bibinfo{year}{2020}\natexlab{}.
\newblock \showarticletitle{Giving voice to silent data: Designing with
  personal music listening history}. In \bibinfo{booktitle}{\emph{Proceedings
  of the 2020 CHI Conference on Human Factors in Computing Systems}}.
  \bibinfo{pages}{1--11}.
\newblock


\bibitem[Wu et~al\mbox{.}(2023)]%
        {wu2023bloomberggpt}
\bibfield{author}{\bibinfo{person}{Shijie Wu}, \bibinfo{person}{Ozan Irsoy},
  \bibinfo{person}{Steven Lu}, \bibinfo{person}{Vadim Dabravolski},
  \bibinfo{person}{Mark Dredze}, \bibinfo{person}{Sebastian Gehrmann},
  \bibinfo{person}{Prabhanjan Kambadur}, \bibinfo{person}{David Rosenberg},
  {and} \bibinfo{person}{Gideon Mann}.} \bibinfo{year}{2023}\natexlab{}.
\newblock \showarticletitle{Bloomberggpt: A large language model for finance}.
\newblock \bibinfo{journal}{\emph{arXiv preprint arXiv:2303.17564}}
  (\bibinfo{year}{2023}).
\newblock


\bibitem[Wu et~al\mbox{.}(2022)]%
        {wu2022ai}
\bibfield{author}{\bibinfo{person}{Tongshuang Wu}, \bibinfo{person}{Michael
  Terry}, {and} \bibinfo{person}{Carrie~Jun Cai}.}
  \bibinfo{year}{2022}\natexlab{}.
\newblock \showarticletitle{Ai chains: Transparent and controllable human-ai
  interaction by chaining large language model prompts}. In
  \bibinfo{booktitle}{\emph{Proceedings of the 2022 CHI conference on human
  factors in computing systems}}. \bibinfo{pages}{1--22}.
\newblock


\bibitem[Xia et~al\mbox{.}(2020)]%
        {xia2020using}
\bibfield{author}{\bibinfo{person}{Meng Xia}, \bibinfo{person}{Yuya Asano},
  \bibinfo{person}{Joseph~Jay Williams}, \bibinfo{person}{Huamin Qu}, {and}
  \bibinfo{person}{Xiaojuan Ma}.} \bibinfo{year}{2020}\natexlab{}.
\newblock \showarticletitle{Using information visualization to promote
  students' reflection on" gaming the system" in online learning}. In
  \bibinfo{booktitle}{\emph{Proceedings of the Seventh ACM Conference on
  Learning@ Scale}}. \bibinfo{pages}{37--49}.
\newblock


\bibitem[Xia et~al\mbox{.}(2019)]%
        {xia2019peerlens}
\bibfield{author}{\bibinfo{person}{Meng Xia}, \bibinfo{person}{Mingfei Sun},
  \bibinfo{person}{Huan Wei}, \bibinfo{person}{Qing Chen},
  \bibinfo{person}{Yong Wang}, \bibinfo{person}{Lei Shi},
  \bibinfo{person}{Huamin Qu}, {and} \bibinfo{person}{Xiaojuan Ma}.}
  \bibinfo{year}{2019}\natexlab{}.
\newblock \showarticletitle{Peerlens: Peer-inspired interactive learning path
  planning in online question pool}. In \bibinfo{booktitle}{\emph{Proceedings
  of the 2019 CHI conference on human factors in computing systems}}.
  \bibinfo{pages}{1--12}.
\newblock


\bibitem[Yang et~al\mbox{.}(2023)]%
        {yang2023pair}
\bibfield{author}{\bibinfo{person}{Kexin~Bella Yang}, \bibinfo{person}{Vanessa
  Echeverria}, \bibinfo{person}{Zijing Lu}, \bibinfo{person}{Hongyu Mao},
  \bibinfo{person}{Kenneth Holstein}, \bibinfo{person}{Nikol Rummel}, {and}
  \bibinfo{person}{Vincent Aleven}.} \bibinfo{year}{2023}\natexlab{}.
\newblock \showarticletitle{Pair-Up: Prototyping Human-AI Co-orchestration of
  Dynamic Transitions between Individual and Collaborative Learning in the
  Classroom}. In \bibinfo{booktitle}{\emph{Proceedings of the 2023 CHI
  Conference on Human Factors in Computing Systems}}. \bibinfo{pages}{1--17}.
\newblock


\bibitem[Yildirim et~al\mbox{.}(2023)]%
        {yildirim2023creating}
\bibfield{author}{\bibinfo{person}{Nur Yildirim}, \bibinfo{person}{Changhoon
  Oh}, \bibinfo{person}{Deniz Sayar}, \bibinfo{person}{Kayla Brand},
  \bibinfo{person}{Supritha Challa}, \bibinfo{person}{Violet Turri},
  \bibinfo{person}{Nina Crosby~Walton}, \bibinfo{person}{Anna~Elise Wong},
  \bibinfo{person}{Jodi Forlizzi}, \bibinfo{person}{James McCann},
  {et~al\mbox{.}}} \bibinfo{year}{2023}\natexlab{}.
\newblock \showarticletitle{Creating Design Resources to Scaffold the Ideation
  of AI Concepts}. In \bibinfo{booktitle}{\emph{Proceedings of the 2023 ACM
  Designing Interactive Systems Conference}}. \bibinfo{pages}{2326--2346}.
\newblock


\bibitem[Zamfirescu-Pereira et~al\mbox{.}(2023)]%
        {zamfirescu2023johnny}
\bibfield{author}{\bibinfo{person}{JD Zamfirescu-Pereira},
  \bibinfo{person}{Richmond~Y Wong}, \bibinfo{person}{Bjoern Hartmann}, {and}
  \bibinfo{person}{Qian Yang}.} \bibinfo{year}{2023}\natexlab{}.
\newblock \showarticletitle{Why Johnny can’t prompt: how non-AI experts try
  (and fail) to design LLM prompts}. In \bibinfo{booktitle}{\emph{Proceedings
  of the 2023 CHI Conference on Human Factors in Computing Systems}}.
  \bibinfo{pages}{1--21}.
\newblock


\bibitem[Zhang et~al\mbox{.}(2023a)]%
        {zhang2023vizprog}
\bibfield{author}{\bibinfo{person}{Ashley~Ge Zhang}, \bibinfo{person}{Yan
  Chen}, {and} \bibinfo{person}{Steve Oney}.} \bibinfo{year}{2023}\natexlab{a}.
\newblock \showarticletitle{VizProg: Identifying Misunderstandings By
  Visualizing Students’ Coding Progress}. In
  \bibinfo{booktitle}{\emph{Proceedings of the 2023 CHI Conference on Human
  Factors in Computing Systems}}. \bibinfo{pages}{1--16}.
\newblock


\bibitem[Zhang et~al\mbox{.}(2023b)]%
        {zhang2023huatuogpt}
\bibfield{author}{\bibinfo{person}{Hongbo Zhang}, \bibinfo{person}{Junying
  Chen}, \bibinfo{person}{Feng Jiang}, \bibinfo{person}{Fei Yu},
  \bibinfo{person}{Zhihong Chen}, \bibinfo{person}{Jianquan Li},
  \bibinfo{person}{Guiming Chen}, \bibinfo{person}{Xiangbo Wu},
  \bibinfo{person}{Zhiyi Zhang}, \bibinfo{person}{Qingying Xiao},
  {et~al\mbox{.}}} \bibinfo{year}{2023}\natexlab{b}.
\newblock \showarticletitle{HuatuoGPT, towards Taming Language Model to Be a
  Doctor}.
\newblock \bibinfo{journal}{\emph{arXiv preprint arXiv:2305.15075}}
  (\bibinfo{year}{2023}).
\newblock


\bibitem[Zheng et~al\mbox{.}(2022)]%
        {zheng2022telling}
\bibfield{author}{\bibinfo{person}{Chengbo Zheng}, \bibinfo{person}{Dakuo
  Wang}, \bibinfo{person}{April~Yi Wang}, {and} \bibinfo{person}{Xiaojuan Ma}.}
  \bibinfo{year}{2022}\natexlab{}.
\newblock \showarticletitle{Telling stories from computational notebooks:
  Ai-assisted presentation slides creation for presenting data science work}.
  In \bibinfo{booktitle}{\emph{Proceedings of the 2022 CHI Conference on Human
  Factors in Computing Systems}}. \bibinfo{pages}{1--20}.
\newblock


\bibitem[Zheng et~al\mbox{.}(2023)]%
        {zheng2023competent}
\bibfield{author}{\bibinfo{person}{Chengbo Zheng}, \bibinfo{person}{Yuheng Wu},
  \bibinfo{person}{Chuhan Shi}, \bibinfo{person}{Shuai Ma},
  \bibinfo{person}{Jiehui Luo}, {and} \bibinfo{person}{Xiaojuan Ma}.}
  \bibinfo{year}{2023}\natexlab{}.
\newblock \showarticletitle{Competent but Rigid: Identifying the Gap in
  Empowering AI to Participate Equally in Group Decision-Making}. In
  \bibinfo{booktitle}{\emph{Proceedings of the 2023 CHI Conference on Human
  Factors in Computing Systems}}. \bibinfo{pages}{1--19}.
\newblock


\bibitem[Zhou et~al\mbox{.}(2023)]%
        {zhou2023comprehensive}
\bibfield{author}{\bibinfo{person}{Ce Zhou}, \bibinfo{person}{Qian Li},
  \bibinfo{person}{Chen Li}, \bibinfo{person}{Jun Yu}, \bibinfo{person}{Yixin
  Liu}, \bibinfo{person}{Guangjing Wang}, \bibinfo{person}{Kai Zhang},
  \bibinfo{person}{Cheng Ji}, \bibinfo{person}{Qiben Yan},
  \bibinfo{person}{Lifang He}, {et~al\mbox{.}}}
  \bibinfo{year}{2023}\natexlab{}.
\newblock \showarticletitle{A comprehensive survey on pretrained foundation
  models: A history from bert to chatgpt}.
\newblock \bibinfo{journal}{\emph{arXiv preprint arXiv:2302.09419}}
  (\bibinfo{year}{2023}).
\newblock


\bibitem[Zimmerman(2002)]%
        {zimmerman2002becoming}
\bibfield{author}{\bibinfo{person}{Barry~J Zimmerman}.}
  \bibinfo{year}{2002}\natexlab{}.
\newblock \showarticletitle{Becoming a self-regulated learner: An overview}.
\newblock \bibinfo{journal}{\emph{Theory into practice}} \bibinfo{volume}{41},
  \bibinfo{number}{2} (\bibinfo{year}{2002}), \bibinfo{pages}{64--70}.
\newblock


\bibitem[Zimmerman and Campillo(2003)]%
        {zimmerman2003motivating}
\bibfield{author}{\bibinfo{person}{Barry~J Zimmerman} {and}
  \bibinfo{person}{Magda Campillo}.} \bibinfo{year}{2003}\natexlab{}.
\newblock \showarticletitle{Motivating self-regulated problem solvers}.
\newblock \bibinfo{journal}{\emph{The psychology of problem solving}}
  \bibinfo{volume}{233262} (\bibinfo{year}{2003}).
\newblock


\end{thebibliography}

\appendix

\end{document}